\documentclass[aps]{revtex4-1} 
\usepackage{xcolor,amssymb}
\usepackage{graphicx}
\usepackage{amsmath,amssymb,bm,epstopdf}
\usepackage{float}
\usepackage{multirow}
\usepackage{array}

\newcommand{\PreserveBackslash}[1]{\let\temp=\\#1\let\\=\temp}
\newcolumntype{C}[1]{>{\PreserveBackslash\centering}p{#1}}
\newcolumntype{R}[1]{>{\PreserveBackslash\raggedleft}p{#1}}
\newcolumntype{L}[1]{>{\PreserveBackslash\raggedright}p{#1}}

\def\u{\bold{u}}
\def\ba#1\ea{\begin{align}#1\end{align}}
\def\bsa#1#2\esa{\begin{subequations}\label{#1}
\begin{align}#2\end{align} \end{subequations}}
\def\lp{\left(}
\def\rp{\right)}
\def\lb{\left[}
\def\rb{\right]}
\def\lcb{\left\{}

\def\NA{\bm{\nabla}}
\def\DEL{\nabla^2}
\def\f{\frac}

\def\p{\partial}

\def\NA{\nabla}

\begin{document}


\title{Shape and size of large-scale vortices : \\ a {generic} fluid pattern in geophysical fluid dynamics}

\author{Louis-Alexandre Couston$^{1,2,*}$, Daniel Lecoanet$^{3}$, Benjamin Favier$^1$, Michael Le Bars$^1$}

\affiliation{$^1$ CNRS, Aix Marseille Univ, Centrale Marseille, IRPHE, Marseille, France \\ $^2$ British Antarctic Survey, Cambridge, CB3 0ET, UK \\ $^3$ Princeton Center for Theoretical Science, Princeton, NJ 08544, USA}





\begin{abstract}

Planetary rotation organizes fluid motions into coherent, long-lived swirls, known as large scale vortices (LSVs), that play an important role in the dynamics and long-term evolution of geophysical and astrophysical fluids. Here, using direct numerical simulations, we show that the shape of LSVs in rapidly-rotating mixed convective and stably-stratified fluids, which approximates the two-layer, turbulent-stratified dynamics of many geophysical and astrophysical fluids, is generic and that their size can be predicted. Specifically, we show that LSVs emerge in the convection zone from upscale energy transfers and can penetrate into the stratified layer. {At the convective-stratified interface, the LSV cores have a positive buoyancy anomaly. Due to the thermal wind constraint, this buoyancy anomaly leads to winds in the stratified layer that decay over a characteristic vertical length scale.} Thus, LSVs take the shape of a depth-invariant cylinder with finite-size radius in the turbulent layer and of a penetrating half dome in the stratified layer. Importantly, we demonstrate that when LSVs penetrate all the way through the stratified layer and reach a boundary that is no slip, they saturate by boundary friction. We provide a prediction for the penetration depth and maximum radius of LSVs as a function of the LSV vorticity, the stratified layer depth and the stratification. Our results, which are applicable for cyclonic LSVs, suggest that while turbulent vortices can penetrate far into the stratified layers of atmospheres and oceans, they should stay confined within the convective layers of Earth's liquid core as well as within those of weakly-rotating stars.

\end{abstract}

\maketitle

\section{Introduction}

Large-scale vortices (LSVs), {i.e. vortices with diameter comparable to or larger than a characteristic geophysical length scale, such as e.g. the depth in shallow fluid layers,  the inner radius in spherical shells or a linear instability horizontal wavelength}, are a key component of geophysical and astrophysical fluids. They are generated by a myriad of processes, ranging from the instability of currents and fronts in oceans \cite{capet2007} to tropical cyclogenesis in the atmosphere  \cite{montgomery2017}. In oceans, LSVs have $O(1-100)$km diameter, weeks to years lifespan  \cite{wang2018,chelton2011}, and they can transport ocean mass, heat and CO$_2$ over long horizontal \cite{dong2014,zhang2014,su2018} and vertical \cite{Klein2009} distances. {LSVs} also influence the background flow \cite{foxkemper2008} and significantly affect plankton productivity and chlorophyll distribution in surface waters  \cite{mahadevan2016,chelton2011b}. 
Planetary atmospheres showcase a wide range of LSVs \cite{heimpel2015}, including long-lived large planetary-scale vortices that control the global circulation and climate (e.g. polar vortex on Earth and Jupiter's Great Red Spot) as well as smaller cyclones with $O(100)$km diameter on Earth \cite{chavas2016} that can have devastating consequences. Earth's outer core, which is made of turbulent liquid iron that powers the {Earth's dynamo}, is also expected to feature numerous LSVs of various sizes \cite{guervilly2019}, such as the high-latitude geomagnetic flux patches \cite{Aurnou2015}, as well as a large-scale north polar vortex  \cite{olson1999}. LSVs are also found in the solar photosphere \cite{brandt1988}, and are expected to exist in accretion disks  \cite{abramowicz1992} and potentially play an important role in planet formation  \cite{marcus2015}. 

LSVs typically result from the breakup of large-scale flows or from upscale energy transfers that feed on small-scale waves and turbulence. In shallow fluid layers that are considered two-dimensional, an inverse cascade guarantees a flux of energy from small scales to LSVs \cite{kraichnan1967,boffetta2012}. However, stars and Earth's outer core can hardly be considered shallow, and LSVs in Earth's atmosphere and oceans often have complicated vertical structures, such that three-dimensional theories are required for realistic predictions. In recent years, it has been shown that rapid rotation enables upscale energy transfers in fully three-dimensional turbulent convection, with a barotropic large vortical mode emerging from the turbulent eddy field \cite{Rubio2014,Favier2014,Guervilly2014,Aurnou2015}. The convection must be turbulent but also strongly constrained by rotation for the LSVs to emerge, a regime known as geostrophic turbulence. Since geostrophic turbulence is common in geophysical and astrophysical fluids, dedicated simulations can unravel the fundamental characteristics of many of the LSVs in nature. Specifically, the relationship between {vortex core pressure anomaly}, maximum velocity and size can be investigated rigorously, and help predict the impact of LSVs not only in the oceanic and atmospheric contexts \cite{chavas2017}, but also in planetary physics and astrophysics. We remark that previous works have focused on simulations of fully-convective fluids with free-slip boundaries \cite{Rubio2014,Favier2014,Guervilly2014}: in this context, no physical process, except for magnetism in a recent study \cite{Maffei2019}, has been found that saturates the growth of LSVs at a natural size, i.e. LSVs always reach the box size, which is unphysical.

We extend previous studies of LSVs in fully-convective fluid systems to LSVs in fluids that are self-organized in a turbulent layer next to a stably-stratified fluid region. Our aim is to investigate the shape and size of a generic model of LSVs similar to eddies in the surface ocean mixed layer penetrating into the thermocline, to cyclones in the Earth's turbulent planetary boundary layer reaching into the upper troposphere and stratosphere, and to LSVs in the convection zone of stars and planetary liquid cores overshooting in adjacent stable (radiative) layers. In the Earth's core context, evidence of a stably-stratified layer at the core-mantle boundary \cite{Buffett2014,Landeau2016a}, or adjacent to the inner core \cite{Alboussiere2010,Hirose2013a}, is recent and has prompted significant interests owing to its potential influence on the geodynamo \cite{Yan2018} and core flows \cite{Vidal2015}. Past studies of penetrating vortices in mixed convective---stably-stratified cores, e.g. \cite{Takehiro2015,Nakagawa2015,Takehiro2018}, are limited to a regime dominated by bulk viscosity that is unlikely to be relevant to planetary dynamics \cite{Aurnou2015}. {Our study solves explicitly turbulent motions, hence may be more directly linked to flows in nature.} Our results may also be applicable to subsurface oceans, e.g. on Enceladus, and subglacial lakes in Antarctica, where a stratified layer can exist close to the bed-water or ice-water boundary due to the nonlinearity of the equation of state for freshwater \cite{thoma2011}. 

Here we show that finite stable fluid layers and boundary friction can control the maximum size of LSVs. This is a result of significant importance since the saturation of upscale energy transfers is not universal but depends on the dissipative or dispersive mechanisms at play at large scales. We demonstrate that the key features of LSVs, including core pressure anomaly, LSV diameter and maximum azimuthal velocity, {can be inferred from the cyclo-geostrophic balance, which is satisfied in the convective region, and that the penetration depth can be inferred from the thermal wind balance, which is approximately satisfied in the stably-stratified region}. We show that the stratification strength and depth of the stable layer control the diameter and extent of penetration of LSVs into the stratified layer, and hence the LSVs' potential to promote vertical exchanges across density interfaces. These two effects are investigated systematically using a suite of direct numerical simulations (DNS) of the Navier Stokes equations with high resolution and long integration time. {We probe the most challenging fully-developed three-dimensional turbulent regimes accessible to such a systematic, exploratory and long-term DNS study, offering novel insights into turbulent LSVs despite being much less turbulent than for natural applications}. We derive an aspect ratio for the penetrating, stably-stratified part of LSVs and deduce an approximate penetration depth and maximum size of LSVs in different geophysical and astrophysical contexts.

\section{Problem formulation}

We investigate the dynamics of a local fluid domain confined between solid top and bottom boundaries and rotating at constant Coriolis frequency $f$ about the vertical $z$ axis  {(cf. figure \ref{fig0})}. We impose fixed temperature conditions, i.e. $T=1$ and {$T=T_{top}<0$ (in dimensionless space), and free slip and either free slip or no slip velocity conditions at the bottom and top boundaries, respectively. {We use a thermal expansion coefficient $\beta(T)$ that is temperature-dependent and changes sign at the inversion temperature $T_{inv}$ (dimensionless), which is smaller than the bottom temperature but larger than the top temperature (cf. equation \eqref{a14} below). As a result, the fluid self organizes into a well-mixed convective layer adjacent to a stably-stratified layer. The use of a nonlinear equation of state with a density maximum is motivated by the similar behavior of water near its density maximum of 4$^\circ C$ \cite{LeBars2015,wang2019,leard2020} and allows to study the coupled dynamics of two-layer mixed turbulent and stably-stratified geophysical and astrophysical fluids. Here, $\beta(T)$ is piecewise-constant and is positive for $T\geq T_{inv}=0$ and negative for $T<T_{inv}=0$ (i.e. we use $T_{inv}$ as reference temperature). Thus, the density is maximum at $T=T_{inv}=0$ and the lower layer is convectively unstable whereas the top layer is stably stratified \cite{Couston2017prf,Couston2018jfm}. {Because we do not prescribe a background hydrostatic state, as is often done in studies of stars \cite{Brummell2002}, in which the convective-stable interface is fixed by imposing a $z$-dependent (rather than $T$-dependent) thermal expansion coefficient, a statistical steady state is achieved only when the two layer depths adjust themself such that the conductive heat flux in the stratified layer equals the convective heat flux in the underlying turbulent layer.} This can take up to a thermal diffusive time, or one tenth of a thermal diffusive time with appropriate initial conditions \cite{Couston2017prf}. This transient, while long, is not restrictive in our simulations since LSV generation relies on slow upscale energy transfers that also take more than or on the order of one tenth of a thermal diffusive time.

We use the {initial} height of the convective layer and the thermal diffusive time as reference length and time scales. The governing equations for velocity $\u=(u,v,w)$, pressure ${p}$, temperature ${T}$ and density anomaly ${\rho}$ are the Navier-Stokes equations in the Boussinesq approximation and can be written in a Cartesian ($x,y,z$) frame of reference and in dimensionless form as
\bsa{a1}\label{a11}
& \p_t \u + \lp\u\cdot\NA\rp\u = - \NA p - PrEk^{-1}\bold{e}_z\times\u + Pr\DEL\u - PrRa \rho \bold{e}_z, \\ \label{a12}
&\p_t T + \lp\u\cdot\NA\rp T = \DEL T, \\ \label{a13}
&\NA\cdot \u = 0, \\ \label{a14}
&\rho=-\beta(T)T=-T\mathsf{H}(T)+ST\mathsf{H}(-T),
\esa
where $\bold{e}_z$ is the upward pointing unit vector, $\mathsf{H}$ is the Heaviside function, $Pr=\nu/\kappa$ is the Prandtl number, $Ra=\beta_sg\Delta d^3/(\nu\kappa)$ is the Rayleigh number, $Ek=\nu/(fd^2)$ is the Ekman number and $S$ is the stiffness parameter, i.e. the ratio of the thermal expansion coefficient in the stable layer to the thermal expansion in the convective layer; $d$ is the initial depth of the convective layer, $\nu$ is the viscosity, $\kappa$ is the thermal diffusivity, $\beta_s$ is the thermal expansion coefficient for the convecting fluid, $g$ is the gravity and $\Delta$ is the temperature difference driving the convection. {We denote} $b=-PrRa\rho$ the buoyancy. Dimensional variables can be recovered from the dimensionless ones using $d$, $d^2/\kappa$, $\Delta$ and $\rho_0\beta_s\Delta $ as characteristic length, time, temperature and density scales, with $\rho_0$ the reference density of the fluid. The control parameters are the horizontal size of the box $L=L_x=L_y$ (in units of initial convective layer depth, $d$), $Ra$, $Ek$, $Pr$, the initial dimensionless stratified layer depth $H$ and the background buoyancy frequency $N=\sqrt{-ST_{top}/H}$ (also known as Brunt-V{\"a}is{\"a}l{\"a} frequency). Here, we set $L=4$ and we select $Pr=1$, $Ra=2\times 10^8$ and $Ek=10^{-5}$ such that the lower convective layer, assuming no effect from the overlaying stratified layer, is in the regime of geostrophic turbulence of fully-convective fluids that feature LSVs \cite{Ecke2014}.

{Given a {total height} $L_z=1+H$, we set $T_{top}$ such that the conductive heat flux through the stratified layer, which is roughly $T_{top}/H$, equals the heat flux in the convective zone, which we estimate from preliminary runs as done in \cite{Couston2018jfm} and \cite{Anders2018}. As a result, the convective zone extends from 0 to $z\approx 1$ and $H$ indicates (approximately) the thickness of the stratified layer both at the initial time and at late times, i.e. once convective motions have set in. We show in figure \ref{fig0} the horizontally-averaged profiles of density and temperature at the initial time (dashed lines) and at a late time (solid lines) for one of our simulations. As expected, there is significant overlap between the initial and final profiles, although it may be noted that the mean interface position $\overline{\xi}$, which is defined as $\overline{T}(z=\overline{\xi})=0$ (overbar denotes horizontal averaging), can in fact be different from its initial position since we set $T_{top}$ based on the convective heat flux at relatively early times in preliminary low-resolution simulations that do not feature LSVs. For the results shown in figure \ref{fig0}, the emergence of LSVs leads to a decrease of the convective heat flux \cite{Guervilly2014} and to a downward movement of the mean interface position, which is shown by the thick solid line in figure \ref{fig0}. We find similar effects of LSVs on the heat flux and similar or smaller downward movement of the mean interface position at the end of the simulations (indicated in table \ref{table}) in all other simulations.} 

We investigate the effect of the buoyancy frequency $N$ (proxy for stratification strength) and of the stable layer depth $H$ on the penetration of turbulent LSVs {(as defined in \S\ref{sec:multi})} into the stratified layer and we demonstrate that LSVs saturate in size when they penetrate through the entire stratified layer and reach a top boundary that is no slip. {We denote simulations with weak, moderate and strong stratification by $\mathcal{W}_H$, $\mathcal{M}_H$ and $\mathcal{S}_H$, respectively, where $H=0.5$, 1 or 2; we denote purely-convective simulations by $\mathcal{C}$, and we use an asterisk to denote simulations with free-slip top boundary conditions (see table \ref{table})}. We run the simulations long enough such that all results presented are at quasi steady-state, i.e. the time- and horizontally-averaged heat flux is constant throughout the depth of the whole fluid and LSV properties are at statistical equilibrium, {i.e. either constant or slowly-varying with time}. Key parameters of the simulations are presented in table \ref{table}, and additional figures are given in the supplementary information (SI) \cite[][]{suppinfo}. {We note that all coherent LSVs in our simulations are cyclonic, in agreement with previous DNS of fully-convective rotating fluids under Boussinesq approximation, which are limited---as is also the case here---to LSVs with relatively large Rossby number, i.e. order 0.1 and above \cite{Favier2014}. Note that for models featuring LSVs with smaller Rossby number, the distribution between cyclonic and anti-cyclonic LSVs is symmetric, i.e. without any bias toward cyclonic vortices \cite{Stellmach2014}.}

\begin{figure}[ht]
\centering
\includegraphics[width=0.6\linewidth]{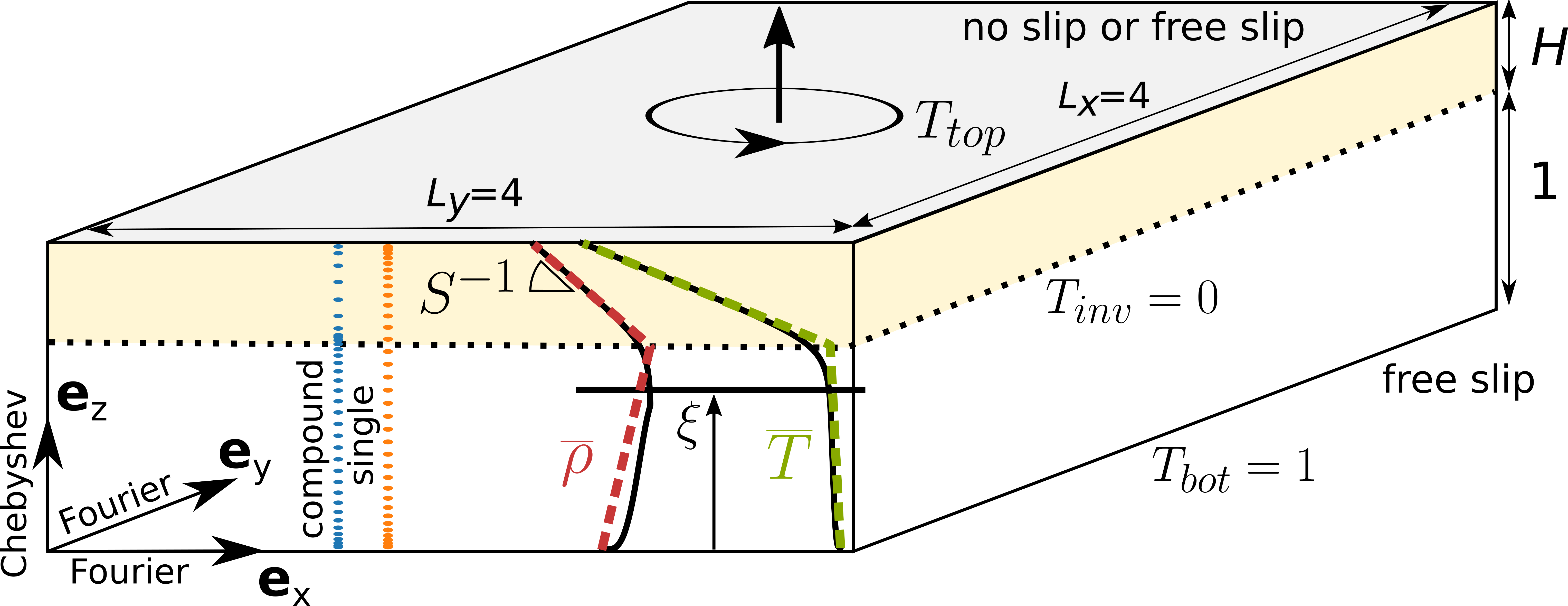}
\caption{{Problem schematic. The fluid is rotating about the $z$ axis and is confined between rigid top and bottom boundaries with fixed temperatures $T_{top}<0$ and $T_{bot}=1$. We use a nonlinear equation of state, such that the fluid is convectively unstable in the lower region (white region) and stably stratified fluid in the upper region (yellow-shaded region). The temperature and density profiles at the initial time (shown by dashed lines for the simulation case $\mathcal{W}_{0.5}$; cf. table \ref{table}) and $T_{top}$ are set such that they overlap with the same profiles at late times (solid lines). As a result, the initial interface position at $z=1$, shown by the horizontal dotted line, gives a reasonable estimate of the mean interface position $\overline{\xi}$ at late times, which is defined as $\overline{T}(z=\overline{\xi})=0$ (overbar denoting horizontal averaging). The location of the grid points in the $z$ direction obtained when using a compound or single Chebyshev basis  are shown by the vertical blue and orange dotted lines, respectively. For the same total spectral resolution (set to 36 modes on this figure), a compound Chebyshev basis (28 and 8 modes in the lower and upper bases, respectively) ensures a high resolution where the two bases are joined, which in our cases is just above the interface between the convective and stably-stratified regions.}}
\label{fig0}
\end{figure}

We solve equations \eqref{a1} using the high-performance, open-source pseudo-spectral simulation code DEDALUS \cite{Burns2018}. We assume horizontal periodicity and we use Fourier modes in the $(x,y)$ directions and Chebyshev modes in the $z$ direction with 3/2 dealiasing. A 2-step implicit/explicit Runge-Kutta scheme is used for time integration. {For simulations $\mathcal{M}_2$ and $\mathcal{S}_1$, we use a compound Chebyshev basis (resolution written as $n_z=n_{z1}[Z]n_{z2}$ in table \ref{table}), i.e. we represent the $z$ dependence using one Chebyshev series from 0 to $Z$ (resolution $n_{z1}$), and a second Chebyshev series from $Z$ to $1+H$ (resolution $n_{z2}$), with added boundary conditions that all variables are continuous at $Z$. A compound basis allows to maintain a high resolution in the convection zone where the Reynolds number based on the rms velocity is order $O(10^3)$, i.e. large (cf. table \ref{table} and note that the Reynolds number based on the maximum velocity is closer to $O(10^4)$), as well as near the interface. Note that $Z$ is the height where the two bases are stitched together and is chosen sufficiently far above the convective-stable interface, i.e. where turbulent motions are weak, such that local grid clustering does not lead to stringent CFL-imposed time-step constraints nor numerical instabilities such as ringing (an example of vertical grid for a single and compound Chebyshev basis is shown in figure \ref{fig0})}. The CFL condition is set to 0.45 and the time step is typically $O(2\times 10^{-7})$. We run the simulations for 0.1 thermal diffusive time or longer, such that each simulation requires on the order of half a million iterations  (cf. resolution and cost details in table \ref{table}). The total computational cost of the present study is approximately 650k cpu-hr.}  \\

\begin{table}\centering \begin{tabular}{C{1cm}R{1cm}R{1cm}R{1cm}R{1cm}R{1cm}R{1cm}R{1cm}R{1cm}R{1cm}R{1cm}C{1.3cm}L{2.8cm}R{0.9cm}}
\hline
Name & $H$ & TBC & $T_{top}$ & $S$ & {$\overline{\xi}$} & {$Re$} & $\mathcal{N}/f$ & $Ro$ & $\ell$ & $\alpha$ & $h$  & $(n_x,n_y,n_z)$ & {cost} \\ 
\hline	 
$\mathcal{W}_{0.5}$ 	& 0.5 	& NS & -24 & 0.1 & 0.71  & 1000	& 0.31 &  0.15 & 0.18 & 6.4 & 2.0  & $(256,256,192)$ & 50k  \\
$\mathcal{W}_{0.5}^{*}$ & 0.5 	& FS & -24 & 0.1 & 0.72  & 3000 	& 0.31 &  0.14 & 0.60 & - & -  & $(256,256,192)$ & 50k  \\
$\mathcal{W}_{1}$ 		& 1 	& NS & -48 & 0.1 & 0.71  & 1300 	& 0.31 &  0.16 & 0.25 & 5.2 & 1.8  & $(256,256,192)$ & 65k \\
$\mathcal{M}_{0.5}$ 	& 0.5 	& NS & -15 & 1 & 0.93  & 2000 	& 0.78 &  0.16 & 0.33 & 1.3 & 0.5  & $(256,256,256)$ & 60k \\
$\mathcal{M}_{1}$ 		& 1 	& NS & -30 & 1 & 0.91  & 2300 	& 0.78 &  0.15 & 0.49 & 1.7 & 0.6  & $(256,256,256)$ & 75k \\
$\mathcal{M}_{2}$ 		& 2 	& NS & -60 & 1 & 0.89  & 2400	& 0.78 &  0.14 & 0.53 & 1.5 & 0.6  & $(256,256,192[1.4]64)$  & 100k \\
$\mathcal{S}_{0.5}$ 	& 0.5 	& NS & -10.5 & 10 & 0.98   & 2100	& 2.05 &  0.12 & 0.61 & 0.4 & 0.2  & $(256,256,256)$ & 60k \\
$\mathcal{S}_{1}$ 		& 1 	& NS & -21 & 10 & 0.98   & 2200	& 2.05 &  0.13 & 0.53 & 0.5 & 0.2  & $(256,256,192[1.1]64)$ & 150k  \\
\hline
$\mathcal{C}$ 			& 0 	& NS & 0 & -  & -  & 380	& -    &  -    &  -  &  -    & -  & $(256,256,128)$ & 20k \\
$\mathcal{C}^*$ 		& 0 	& FS & 0 & -  & -  & 2200	& -    &  0.11 & 0.52 &  -    & -  & $(256,256,128)$ & 20k \\
\hline
\end{tabular}\vspace{-0.in}\caption{Key input and output parameters of the simulations. $\mathcal{W}$, $\mathcal{M}$ and $\mathcal{S}$ denote simulations with weak, moderate and strong stratification, and the subscript indicates the depth of the stratified layer $H$ relative to the convective layer height. TBC and $T_{top}$ indicate the type of velocity condition and the imposed temperature on the top boundary (NS=no slip; FS=free slip); simulations with a free-slip top boundary have a star superscript. $S$ is the stiffness parameter, {$\overline{\xi}$ is the mean interface position at the end of the simulations}, {$Re$ is the Reynolds number based on the rms velocity in the middle of the convection zone ($z=0.5$) at $t=0.1$ (note that $Re$ is still slowly varying in some cases at that time)} and $\mathcal{N}/f=EkN/Pr$ is the normalized buoyancy frequency. $Ro$ is the Rossby number, $\ell$ is the radius of maximum velocity, $h$ is the \textit{e}-folding velocity decay height of the dominant LSV, and $\alpha=h/\ell$. $(n_x,n_y,n_z)$ are the number of Fourier and Chebyshev modes (before 3/2 dealiasing) in the $(x,y,z)$ directions, and cost is the approximate computational cost in cpu-hr {(see the text for more details)}. $\mathcal{C}$ and $\mathcal{C}^*$ refer to fully-convective simulations, i.e. without a stably-stratified layer.}\label{table}\end{table}

\section{Results}

\subsection{Importance of stably-stratified layers}
We show in figure \ref{fig1} snapshots of the horizontal velocity $V=\sqrt{u^2+v^2}$ as well as of the temperature field at a late time for two fully-convective simulations (figures \ref{fig1}A$^*$,A) and two convective---stably-stratified simulations (figures \ref{fig1}B,C). In fully-convective simulations, i.e.  without a stratified layer, a LSV emerges when the top boundary is free-slip (figure \ref{fig1}A$^*$), but not when the top boundary is no-slip (cf. figure \ref{fig1}A). This indicates, in agreement with previous studies \cite{kunnen2016}, that boundary friction inhibits upscale energy transfers in fully-convective fluids such that large-scale barotropic vortices cannot be obtained in current DNS (i.e. which are limited to relatively low Reynolds number) with no-slip boundaries {(except for a recent work, cf. \cite{guzman2020})}. With a stratified layer ($H> 0$), we find that one or several LSVs always emerge for the same convective parameters as in figure \ref{fig1}A, even with a no-slip top boundary (cf. one LSV in figure \ref{fig1}B and several smaller LSVs in figure \ref{fig1}C). This means that stratified layers {shield} upscale energy transfers and LSVs against boundary friction, which is a fundamental and important result for planetary cores and potentially for Earth's oceans and subsurfaces oceans of icy moons: subadiabatic layers of planetary cores and oceans' pycnoclines may {shield} LSVs against boundary friction at e.g. the core-mantle boundary or the seabed. We note that LSVs are expected to emerge despite no-slip boundaries in reduced models of fully-convective fluids assuming asymptotically-large rotation and turbulence intensity \cite{Plumley2017}. Therefore, a stratified layer mitigating boundary friction may not be always necessary to explain the emergence of LSVs, though it should broaden their domain of existence to cases accessible to DNS and possibly laboratory experiments \cite{Cheng2015}. We recall that the bottom boundary is free slip in all our simulations since LSVs do not emerge in a convective fluid directly adjacent to a no-slip bottom boundary for our choice of parameters.  \\

\begin{figure}[ht]
\centering
\includegraphics[width=0.85\linewidth]{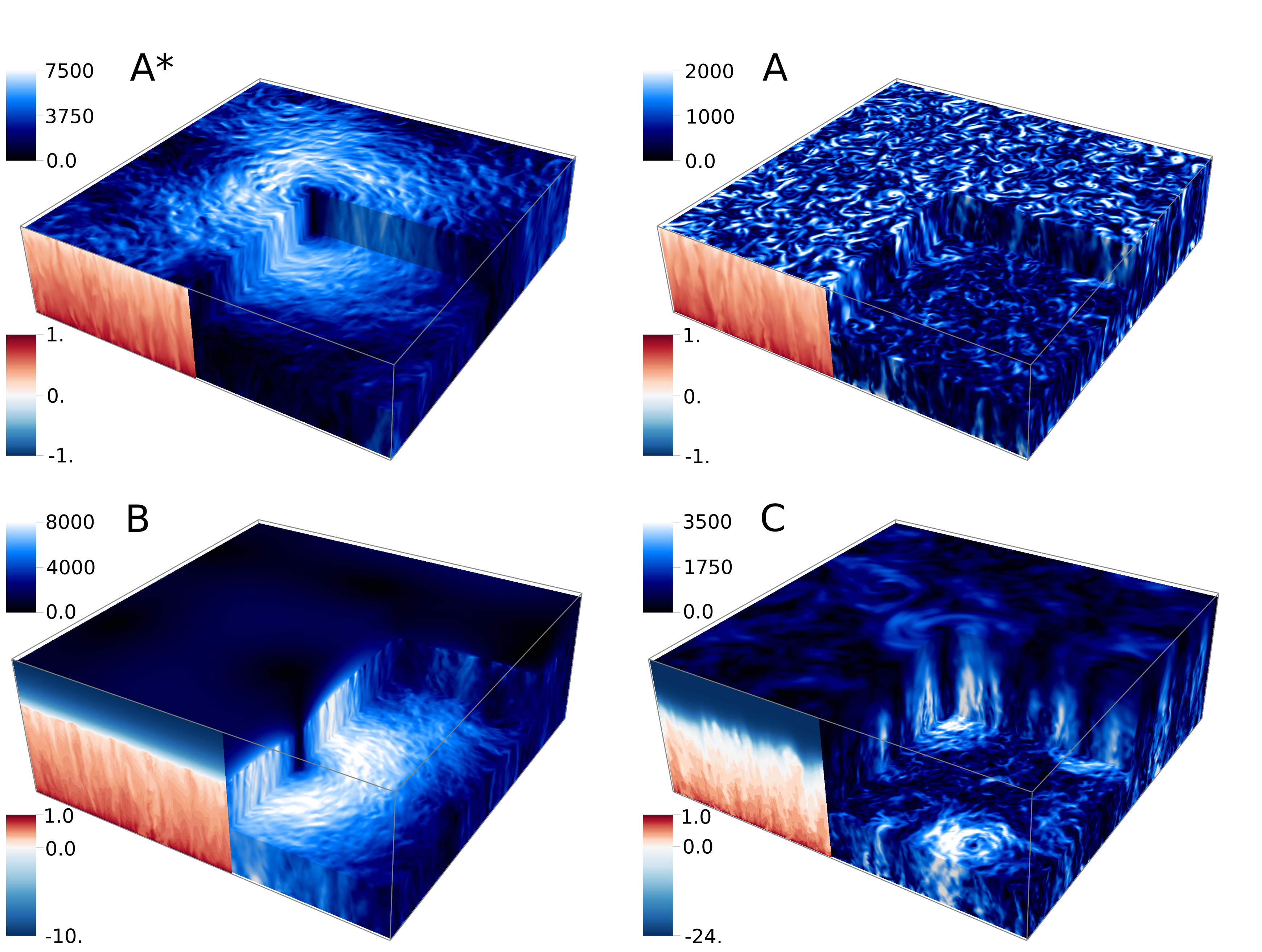}
\caption{{Snapshots of the horizontal velocity amplitude $V$ (blue colormap in three dimensions) and of the temperature field $T$ (red-blue colormap on vertical slice) in purely-convective simulations and in  mixed convective---stably-stratified fluid simulations. The results are shown  at statistical steady state for simulations (A$^*$) $\mathcal{C}^*$, (A) $\mathcal{C}$, (B) $\mathcal{S}_{0.5}$ and (C) $\mathcal{W}_{0.5}$ of table \ref{table}. Blue (resp. red) colors of the temperature field highlight the stratified (resp. turbulent) fluid region. The velocity field on the upper panel is shown at $z=L_z-0.04$, i.e. slightly below the top boundary where velocities are zero with no-slip conditions. In (B) the LSV is wide and weakly-penetrating while in (C) there are several tall LSVs that penetrate far into the stratified fluid.}}
\label{fig1}
\end{figure}
\subsection{Horizontal saturation}
%
 LSVs in nature grow and saturate at a finite size either because there is a physical mechanism that prevents their growth beyond a certain point or because they reach the boundaries of the geophysical or astrophysical fluid domain. Previous studies of fully-convective Cartesian fluid domains with free-slip boundaries have always reported LSVs growing to the box size \cite{Favier2014,Guervilly2014}. This is a severe limitation to the application of existing numerical local models to natural cases, since the box size in periodic simulations is not a real physical quantity. Here, we demonstrate that boundary friction through a stably-stratified layer provides a natural saturation mechanism for LSVs, and that the final \textit{natural} diameter depends on the stratification strength $N$ and depth $H$ of the stable layer. Figures \ref{fig1}B,C clearly show the sensitivity of the natural diameter of LSVs with $N$ (all other parameters being the same). In figure \ref{fig1}B the stratification is strong and the LSV fills up the entire domain, suggesting that the natural LSV diameter is large, larger in fact than the horizontal extent of the domain. In figure \ref{fig1}C, on the other hand, the stratification is weak and several LSVs co-exist, merge and split but on average do not grow bigger than about a third of the domain size, suggesting that the LSVs saturate naturally at a moderate diameter and do not experience numerical confinement.

In order to assess which simulations feature domain-filling LSVs (i.e. confined numerically) and which simulations feature LSVs saturating naturally, we show in figure \ref{fig2}A the integral length scale at mid-depth of the convective layer, i.e. $L_0(z=0.5)$, which is a proxy for LSV diameter {(note that $L_0$ is invariant with depth in the convective layer; cf. figure \ref{fig2}E)}. The integral length scale is given by
\ba{}\label{L0}
L_0(z) = \f{\int \lp |\hat{u}|^2+|\hat{v}|^2 \rp k^{-1} dk}{\int \lp |\hat{u}|^2+|\hat{v}|^2 \rp dk},
\ea
with $k=(k_x^2+k_y^2)^{1/2}$ the horizontal wavenumber (assuming horizontal isotropy) and a hat denotes Fourier transform in $(x,y)$. {In all cases, we see that $L_0(z=0.5)$ first grows rapidly with time, then slows down and eventually reaches a mean value by $t\approx 0.1$, which is constant or slightly increasing (recall that $t$ is normalized by the long thermal diffusive time scale)}. The fully-convective simulation $\mathcal{C}$ has $L_0(z=0.5) \approx 0.5 \ll L=4$ at saturation (solid black line), which is much smaller than $L_0(z=0.5)$ in all other cases. This is because there is no LSV in $\mathcal{C}$ due to the no-slip condition, as seen in figure \ref{fig1}A. Conversely, for simulation $\mathcal{C}^*$ (black line with crosses), $L_0(z=0.5) \approx 3.3$ {at the final time (still slightly increasing)} and is roughly the maximum attainable since in this case the cyclone saturates at the size of the numerical domain $L=4$. With a stratified layer and a no-slip top boundary (colored solid lines), we find that $L_0(z=0.5)$ at steady-state  increases with the stratification strength $N$ (blue to orange to green) and with the stable layer depth $H$ (thin to thick lines). The simulations with the strongest stratification ($\mathcal{S}_{0.5}$ and $\mathcal{S}_1$) and with the thickest stratified layer ($\mathcal{M}_2$) have $L_0(z=0.5) \approx 3.3$, i.e. feature a unique LSV that has reached the domain size. The effect of the stratified layer depth on the number and diameter of LSVs in simulations with moderate stratification can be seen in figure \ref{fig2}B-D where we show the horizontal velocity $V$ in the middle of the convection zone at $t\approx 0.1$: clearly, LSVs saturate at smaller sizes when the stratified layer becomes shallower (figure \ref{fig2}B to \ref{fig2}D). It is worth noting that LSVs saturate  naturally only when the top boundary is no slip and provides friction. Indeed, while $L_0(z=0.5)\approx 1.2$ for $\mathcal{W}_{0.5}$ with no slip, $L_0(z=0.5)\approx 3.3$ for $\mathcal{W}_{0.5}^*$ with free slip and the LSV fills up the entire domain (i.e. saturates numerically). In fully-convective systems it has been shown that the box-filling LSVs can be replaced with large-scale jets when the domain aspect ratio is changed \cite{Guervilly2017a,Julien2018}. Our results suggest that moderate-size, penetrating LSVs should be robust against such changes since they saturate at diameters smaller than the horizontal extent of the numerical domain. {We show in figure \ref{fig2}E the integral length scale $L_0(z)$ as a function of depth for simulations $\mathcal{M}_{0.5}$, $\mathcal{M}_{1}$ and $\mathcal{M}_{2}$. $L_0$ is depth-invariant within the convective bulk that extends up to $z\approx 1$, and then increases slightly with height because small-scale motions are inhibited in the stratified layer. Similar results for $L_0(z)$ as a function of $z$ are obtained for other simulations and shown in figure 1 in SI \cite[][]{suppinfo}.}  \\

\begin{figure}[ht]
\centering
\includegraphics[width=0.85\linewidth]{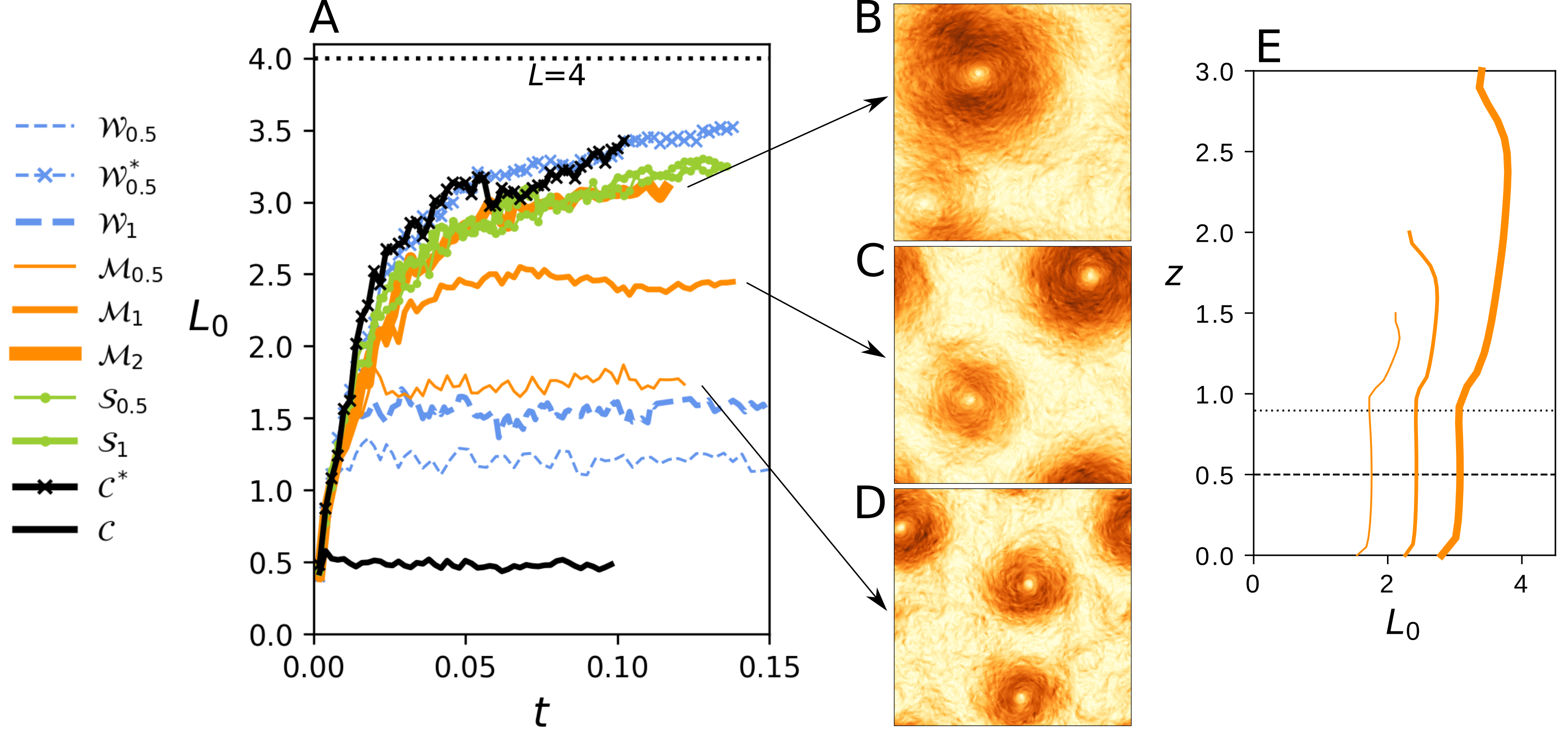}
\caption{(A) Integral length scale $L_0$ (proxy for LSV diameter) in the middle of the convection zone at $z=0.5$ as a function of time $t$ for the simulations of table \ref{table}. {Blue dash-dash, orange solid and green solid-dot lines} denote weak, moderate, strong stratifications, and thicker lines correspond to thicker stable fluid layers. {Crosses} indicate results obtained with a free-slip condition on the top boundary and the dotted line shows the box size $L=4$. (B-D) Snapshots of horizontal velocity $V$ in the middle of the convection zone at steady-state for simulations $\mathcal{M}_{2}$, $\mathcal{M}_{1}$, $\mathcal{M}_{0.5}$. {(E) Depth dependence of $L_0$ for $\mathcal{M}_{2}$ (thick solid line), $\mathcal{M}_{1}$ (regular solid line) and $\mathcal{M}_{0.5}$ (thin solid line) averaged over a 0.02 time window at the end of the simulations. The dashed line shows $z=0.5$ while the dotted line indicates $z=0.91$, which is approximately the mean interface position of the three simulations (cf. table \ref{table}).}}
\label{fig2}
\end{figure}

\subsection{Multiple LSVs}\label{sec:multi}

{The instantaneous velocity field, as shown in figure \ref{fig1} and figures \ref{fig2}B-D, exhibits significant variability due to  turbulent motions that make the identification of LSVs difficult. Here, in an effort to identify and quantify the number of LSVs in our simulations, we propose an unambiguous---although somewhat arbitrary---definition of LSVs based on the smoother pressure field. We recall that LSVs are cyclonic in our simulations, hence correspond to low-pressure systems. Let us define the pressure anomaly $p'$ as $p'=p-\overline{p}$, where overbar denotes the horizontal average. We draw  contours of depth-averaged pressure anomaly satisfying $\int_0^1 p' dz = \min(\int_0^1 p' dz)/3$ and count each closed contour as an LSV if the longest dimension of the contour equal or exceed half the convective-layer depth. Thus, all LSVs have a much larger diameter than the critical wavelength of rapidly-rotating convection, which is $\lambda_c \approx 2\pi/(1.3Ek^{-1/3}) \approx 0.1$ for $Ek=10^{-5}$ \cite{Chandrasekhar1961}. Figures \ref{fig2b}A-C show the results of our detection algorithm based on the pressure field at a late time for simulations $\mathcal{M}_{0.5}$, $\mathcal{M}_{1}$ and $\mathcal{M}_{2}$: all closed contours in these figures count as LSVs because their longest dimension is larger than 0.5. Figure \ref{fig2b}D shows the number of LSVs detected using this algorithm as a function of time. All simulations have at least one LSV,  except fully-convective simulation $\mathcal{C}$, which we know does not feature LSVs and is not shown. In agreement with the results shown in figure \ref{fig2}A, the simulations with weak stratification (blue lines) or moderate stratification but small $H$ (thin orange line), have multiple LSVs. Simulation $\mathcal{W}_{0.5}$ (thin dashed blue line) has between 1 and 5 LSVs at any one time, 5 being the most number of LSVs ever obtained. For completeness, we show in figure \ref{fig2b}E the minimum of pressure anomaly as a function of time. The minimum of pressure anomaly for simulation $\mathcal{C}$ remains much larger than the minimum of pressure anomaly for all other simulations, especially, past the initial transients, when the difference is about a factor 10. We conclude that the detection of LSVs based on the minimum of pressure anomaly is appropriate, although one could in addition enforce a condition on the minimum of pressure anomaly being smaller than a threshold value, e.g. in our case $\min(\int_0^1 p' dz)<-10^8$ (horizontal dashed line in figure \ref{fig2b}E). It may be noted that the minimum of pressure anomaly still decreases with time for the 5 simulations with a single LSV. This occurs because box-filling LSVs continue to slowly intensify after reaching the box size over time scales that exceed our simulation time.}

\begin{figure}[ht]
\centering
\includegraphics[width=0.9\linewidth]{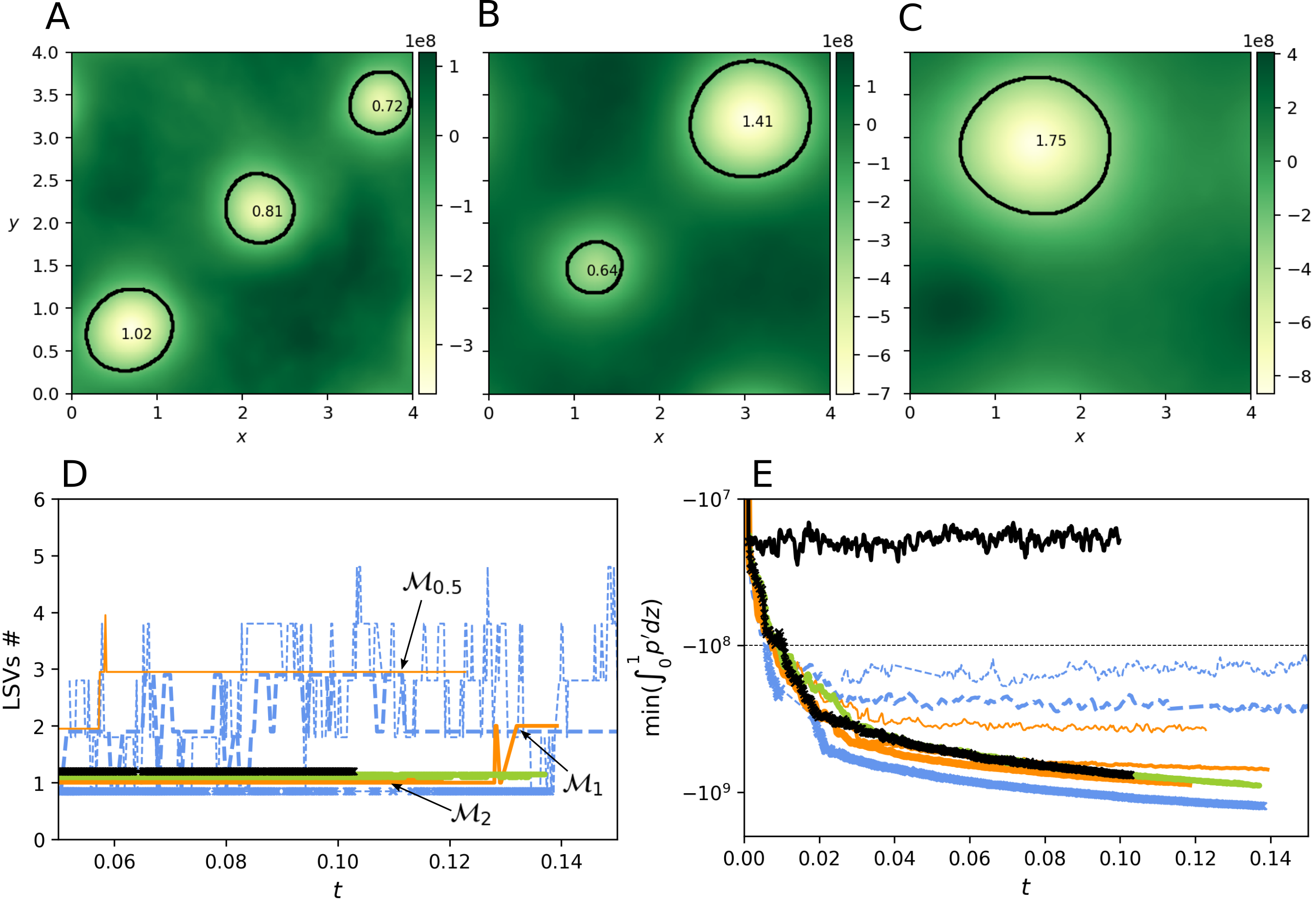}
\caption{{Illustration of our pressure-based detection algorithm of large-scale vortices. Figures \ref{fig2b}A-C show the instantaneous pressure field $p'$ for simulations $\mathcal{M}_{0.5}$, $\mathcal{M}_{1}$ and $\mathcal{M}_{2}$. The solid lines are contours of depth-integrated pressure anomaly and highlight cyclonic LSVs, which are low-pressure systems (see the text for more details); the inside numbers indicate the longest dimension of the contours. Figure \ref{fig2b}D-E show the number of LSVs detected and the minimum of pressure anomaly integrated from $z=0$ to $z=1$, respectively, as functions of time and with line styles denoting the same simulations as in figure \ref{fig2}A. Note that the number of LSVs is always an integer, though we offset some of the results in figure \ref{fig2b}D by a small number in order to see all lines; $\mathcal{M}_2$ (difficult to see) has only one LSV at all times.}}
\label{fig2b}
\end{figure}

\subsection{Shape of penetrating LSVs}\label{subsec:shape}
%
In order to understand why weak stratification and small stratified layer depth  (resp. strong stratification and large stratified layer depth) lead to small (large) LSV diameters, we now provide a phenomenological description of the axisymmetric shape of LSVs. 

{We first demonstrate that LSVs satisfy the cyclo-geostrophic and hydrostatic equations. Let us identify the centre $(x_c,y_c)$ of the dominant LSV in each simulation from the minimum of pressure anomaly integrated from $z=0$ to $z=1$. The cyclo-geostrophic and hydrostatic equations can be written in a cylindrical ($r,\theta,z$) coordinate system centred on $(x_c,y_c)$ as
\bsa{geo2}\label{geo21}
&\p_r \overline{p}  = \f{Pr\overline{v_{\theta}}}{Ek}+\f{\overline{v_{\theta}}^2}{r}, \\ \label{geo22}
&\p_z \overline{p} = \overline{b}, 
\esa
with $\overline{v_{\theta}}$ the azimuthal velocity and overbar now denotes azimuthal and time averaging over a $\sim 0.02$ time period. Note that we consider time-averaged variables for simplicity from here onward since instantaneous profiles of buoyancy and pressure terms---while they follow reasonably well the trends of the time mean---display fluctuations that make the analysis difficult (cf. figure 2 in SI \cite[][]{suppinfo}). Following \cite{Hassanzadeh2012} we expand the pressure and buoyancy variables as $\overline{p}(r,z)=\overline{p}'(r,z)+\overline{p}_{\infty}(z)$ and $\overline{b}(r,z)=\overline{b}'(r,z)+\overline{b}_{\infty}(z)$ with subscript $\infty$ denoting the far-field value. Equations \eqref{geo2} then require $\p_z\overline{p}_{\infty}=\overline{b}_{\infty}$ and can be rewritten for the anomalous variables as (cf. more details in appendix \ref{appA})
\bsa{geo3}\label{geo31}
&\p_r \overline{p}'  = \f{Pr\overline{v_{\theta}}}{Ek}+\f{\overline{v_{\theta}}^2}{r}, \\ \label{geo32}
&\p_z \overline{p}' = \overline{b}'. 
\esa 
We show in figure \ref{fgeo}(A-B) (resp. (C-D)) the terms on the left-hand-side and on the right-hand-side of equation \eqref{geo31} (resp. equation \eqref{geo32}) by solid lines and circles, respectively, at four different heights, for simulations $\mathcal{W}_{0.5}$ and $\mathcal{M}_1$ (similar results are obtained and shown in figure 2 in SI for other simulations \cite[][]{suppinfo}). There is an excellent overlap between the solid lines and the circles at all heights, i.e. both above and below the convective-stratified interface, which means that the cyclo-geostrophic equation (figure \ref{fgeo}(A-B)) and the hydrostatic equation (figure \ref{fgeo}(C-D)) are satisfied in both the turbulent bulk and the stratified layer. The terms of the cyclo-geostrophic equation are much larger than the terms of the hydrostatic equation for $z<1$, such that the flow satisfies cyclo-geostrophic balance at leading order in the convective layer (i.e. the flow is mostly barotropic). For $z\geq 1$, all terms of equations \eqref{geo3} have the same magnitude, such that the flow is controlled by both cyclo-geostrophic and hydrostatic balances in the stratified layer. The dependence of the azimuthal velocity with height in the stratified layer where hydrostatic balance becomes important is of particular interest and can be inferred from equations \eqref{geo3}. Combining the $z$ derivative of equation \eqref{geo31} and the $r$ derivative of equation \eqref{geo32}, we obtain
\ba{}\label{geo0}
\p_z \lp \overline{v_{\theta}} + \f{Ek\overline{v_{\theta}}^2}{Pr r} \rp  = \f{Ek}{Pr} \p_r \overline{b}'.
\ea
Assuming small azimuthal velocities, i.e. $\f{Ek\overline{v_{\theta}}}{rPr} \ll 1$, equation \eqref{geo0} then simplifies into the thermal wind equation, i.e.
\ba{}\label{geo}
\p_z  \overline{v_{\theta}} = \f{Ek}{Pr} \p_r \overline{b}',
\ea
which we use to relate the vertical variations of $\overline{v_{\theta}}$ to the radial derivative of $\overline{b}'$ in our phenomenological description of LSVs below. Equation \eqref{geo} is not as accurate as equation \eqref{geo0}; however, we have checked that geostrophic balance holds relatively well at all heights in our simulations (cf. dashed blue lines overlapping with circles in figure \ref{fgeo} and in figure 2 in SI \cite[][]{suppinfo}) such that equation \eqref{geo} is sufficient to describe the phenomenology of the flow. We denote by  $\overline{\zeta}$ the vertical vorticity; $\overline{\zeta}$ is related to the velocity via $r\overline{\zeta}=\p_r (\overline{v_{\theta}}r)$, such that $\p_z\overline{\zeta}$ and $\p_z\overline{v_{\theta}}$ have generally the same sign in a LSV.}

\begin{figure}
\centering
\includegraphics[width=0.55\textwidth]{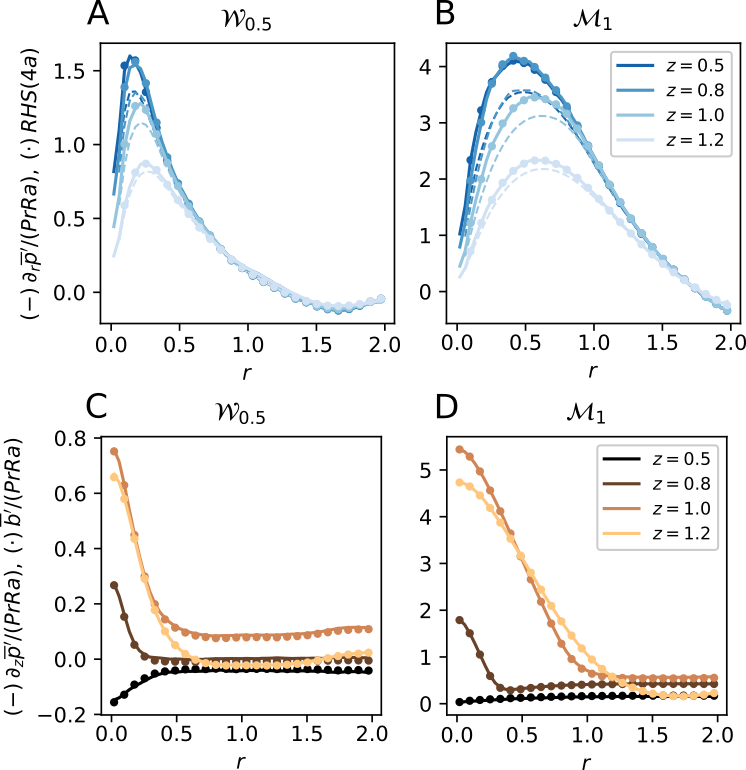}
\caption{{(A-B) Cyclo-geostrophic and (C-D) hydrostatic balance. (A-B) The blue solid lines and the blue circles show the left-hand-side (pressure term) and right-hand-side (velocity term) of the cyclo-geostrophic equation \eqref{geo31}, respectively. For completeness we also show the first (geostrophic) term on the right-hand-side of equation \eqref{geo31} divided by $PrRa$ with dashed lines. (C-D) The solid lines and the brown circles show the left-hand-side (pressure term) and right-hand-side (buoyancy term) of the hydrostatic equation \eqref{geo32}. The good agreement between solid lines and circles is shown as a function of $r$ for $z=0.5,0.8,1.0,1.2$ for simulations $\mathcal{W}_{0.5}$ (A,C) and $\mathcal{M}_{1}$ (B,D) of table \ref{table}.}}\label{fgeo}
\end{figure}

We now show in figure \ref{fig3}A a schematic of the axisymmetric structure of LSVs in mixed convective and stably-stratified fluids. The schematic is based on the vertical vorticity field $Ek\overline{\zeta}/Pr$ {(which is normalized such that it corresponds to the local Rossby number)} obtained in DNS and shown in figures \ref{fig3}B,C for simulations $\mathcal{M}_2$ and $\mathcal{M}_{0.5}$ (similar results are obtained for all simulations and shown in figure 2 in SI \cite[][]{suppinfo}). We denote $\ell$ the radius of maximum azimuthal velocity at the base of the stratified vortex cap and $h$ the \textit{e}-folding decay height of azimuthal velocity in the stratified layer (penetration depth for short). {It may be noted that $\ell$ is approximately 4 times smaller (or more) than the integral length scale $L_0$ shown in figure \ref{fig2}A, which is because $\ell$ is a measure of the radius of the cyclone, i.e. half the diameter, while $L_0$ is the full extent (wavelength) of a cyclone---anti-cyclone pair (where the anti-cyclone is not coherent, i.e. but a region of negative vorticity)}. The black dashed line shows the interface between the convection zone and the stably-stratified layer.
We show the LSV in figure \ref{fig3}A as a cylinder of depth-invariant vorticity within the convective layer and as a half dome of vertically-decaying vorticity in the stably-stratified fluid, which we call the stratified vortex cap. The large vorticity inside the LSV inhibits turbulence compared to the outside in the convective layer \cite{Favier2019}, such that there is less and less mixing toward the LSV centre. This results in a vertical temperature gradient steepest at the LSV centre \cite{Guervilly2014} and, accordingly, a downward depression of the isothermal of maximum density (black dashed line) also toward the LSV centre. We denote by $\delta$ the amount by which the stratified vortex cap sinks into the convective zone, called the restratification depth, in reference to the restratification of the oceanic surface layer due to eddies \cite{Klein2008}. The decrease (in magnitude) of the vertical temperature gradient with radius results in a negative temperature anomaly, $\overline{T}'=\overline{T}-\overline{T}_{\infty}$, in the LSV centre. This anomaly is shown by the light red-coloured cone in figure \ref{fig3}A and is small, as is the buoyancy anomaly $\overline{b}'=\overline{T}'<0$, in most of the convective layer. As a result, the LSV roughly satisfies the Taylor-Proudman theorem, i.e. is depth-invariant, in the convective layer (cf. equation \eqref{geo}). The negative temperature anomaly increases with height, such that at and above the base of the stably-stratified layer, it translates into a positive and potentially large buoyancy anomaly $\overline{b}'=-S\overline{T}'$. This positive buoyancy anomaly drives the decay of the azimuthal velocity with height above the black dashed line according to the thermal wind balance, i.e. equation \eqref{geo}, which is why the stratified LSV has a half-dome shape. When $S$ increases, i.e. the stratification becomes stronger, $\overline{b}'$ increases and so does $\p_r \overline{b}'$, such that the aspect ratio $h/\ell$ of a LSV must decrease in order to satisfy the thermal wind balance. This explains why in a strongly-stratified fluid LSVs appear as wide weakly-penetrating columns  (cf. figure \ref{fig1}B), while in a weakly-stratified fluid they appear as tall narrow columns (figure \ref{fig1}C).

\begin{figure}
\centering
\includegraphics[width=0.85\linewidth]{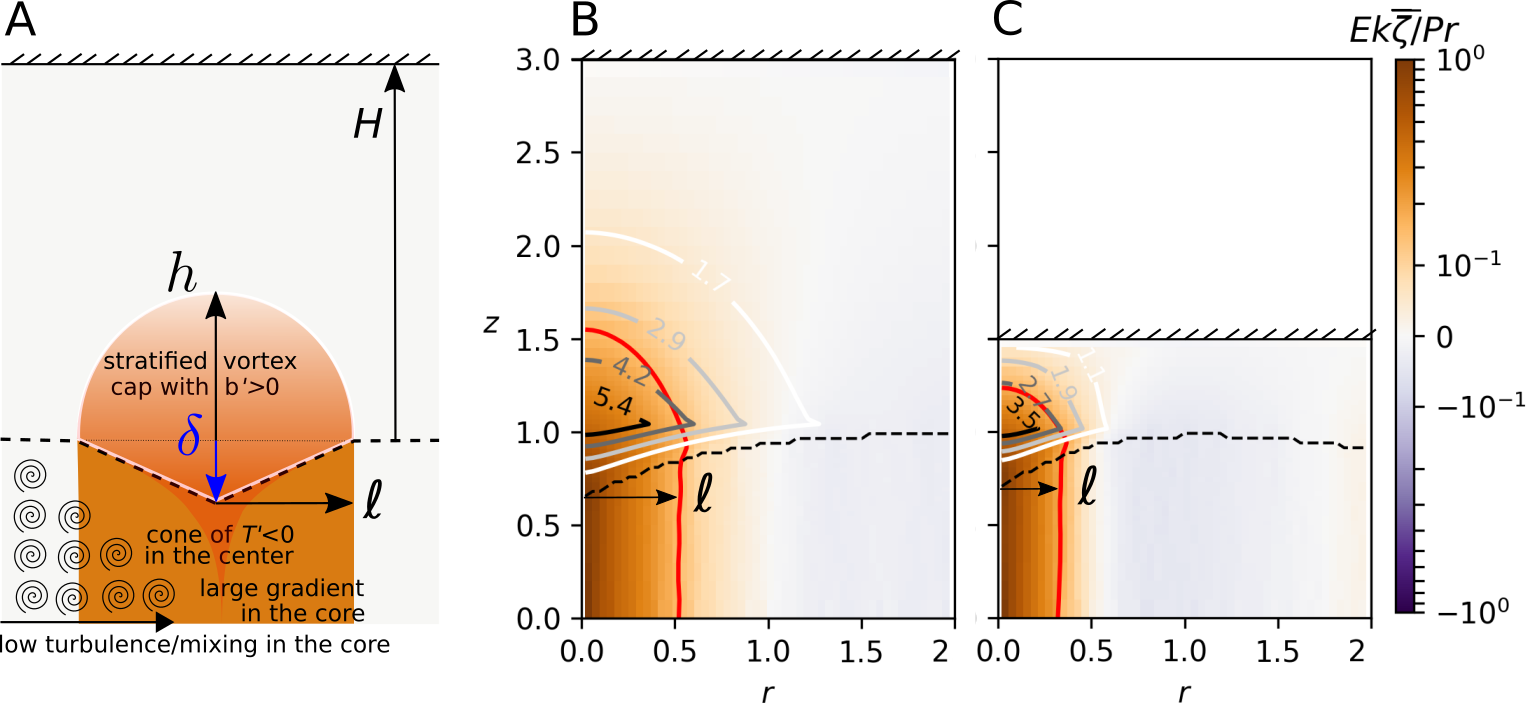}
\caption{(A) Schematic of the axisymmetric structure of LSVs obtained in DNS with $\ell$, $h$ and $\delta$ the LSV diameter, penetration depth and restratification depth. The stratified vortex cap is the part of the LSV that is above the convective-stable interface (black dashed line) and is highlighted by a solid white line. The red cone highlights the region where the temperature anomaly $\overline{T}'<0$. (B-C) Map of vertical vorticity $Ek\overline{\zeta}/Pr$ in a cylindrical coordinates system centred on the vortex core after time and azimuthal averaging for simulations $\mathcal{M}_2$ and $\mathcal{M}_{0.5}$, respectively (cf. table \ref{table}).  {We multiply $\overline{\zeta}$ by $Ek/Pr$ such that the variable shown corresponds to the local Rossby number, which is always order 1 in the vortex core and smaller than 1 away from the core}. The solid lines with grey color scale show isocontours of buoyancy anomaly $\overline{b}'>0$. The red solid line is a contour of constant vorticity.}
\label{fig3}
\end{figure}

Figures \ref{fig3}B,C show the vertical vorticity for simulations $\mathcal{M}_2$ (figure \ref{fig3}B) and $\mathcal{M}_{0.5}$ (figure \ref{fig3}C), i.e. which have a deep and shallow stratified layer, respectively, but same parameters otherwise. As described above, the stratified vortex cap has a positive buoyancy anomaly in both cases (as shown by the gray contours), which is balanced by a vorticity decay with height above the convective-stable interface (dashed line). However, while the penetration of the vortex cap is small compared to the stratified layer thickness $H$ in figure \ref{fig3}B, the penetration depth is large enough compared to $H$ in figure \ref{fig3}C such that the LSV is confined vertically. The maximum vorticity does not change significantly between the two simulations and the buoyancy anomaly is smaller in figure \ref{fig3}C than in figure \ref{fig3}B (cf. in-line numbers). Thus, $|\p_z\overline{v_{\theta}}|$ is larger for a vertically-confined LSV than for a vertically-unconfined LSV, which means that confined LSVs must decrease in diameter (compared to their unconfined counterparts), i.e. such that $|\p_r\overline{b}'|$ increases, in order to maintain thermal wind balance. As a result, boundary friction makes the LSVs saturate naturally in general and in particular in figure \ref{fig3}C, because it imposes a sharp vorticity decay that can only be balanced by a reduction of the LSV diameter. It can be noted that the horizontal narrowing of vertically-confined LSVs does not apply when the top boundary is free-slip since in this case the vorticity doesn't decay faster than when it is unconfined. \\

\subsection{Aspect ratio of the stratified vortex cap}\label{sec4}

\begin{figure}
\centering
\includegraphics[width=0.7\linewidth]{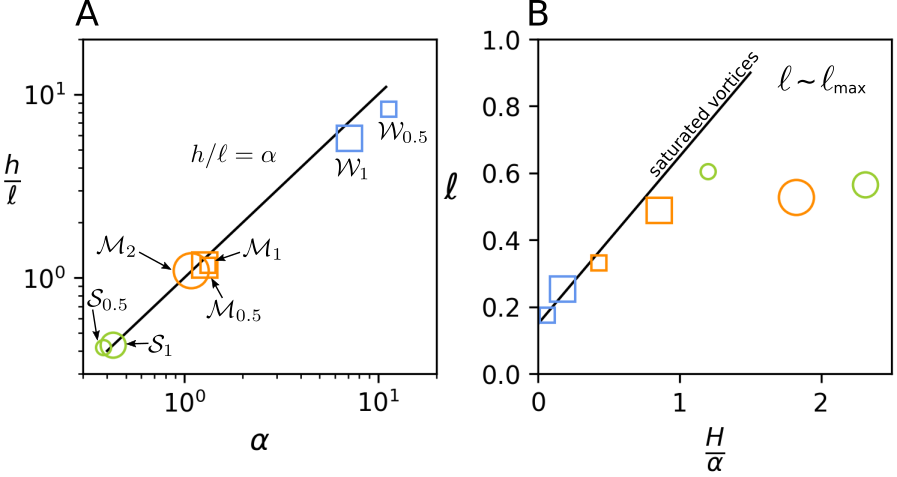}
\vspace{-0.1in}\caption{(A) Aspect ratio $h/\ell$ of the stratified vortex cap against the theoretical prediction \eqref{alpha2} for $\alpha$. {Blue, orange, green colors highlight simulation results with weak, moderate and strong stratification (as in figure \ref{fig2}), and the size of the symbols is proportional to the thickness of the stable layer, i.e. large symbols highlight simulations with thick stratified layers. Circles denote simulations with LSVs saturating at the box size, while squares highlight simulations with LSVs saturating naturally, i.e. from vertical confinement.} (B) LSV radius $\ell$ as a function of $H/\alpha$. The solid line shows that the radius of saturated LSVs follows the same trend as the maximum radius $\ell_{max}=H/(2\alpha)$ predicted for LSVs that are confined vertically. LSVs that are not confined vertically have $\ell<\ell_{max}$ and saturate at the box size (cf. three rightmost symbols shown as circles).}
\label{fig4}
\end{figure}
%
The aspect ratio of the stratified vortex cap, $\alpha=h/\ell$, is a function of the normalized stratification strength $\mathcal{N}/f$, with $\mathcal{N}=fEkN/Pr$ the dimensional buoyancy frequency, and the Rossby number of the LSV, i.e. $Ro=Ek(v_{\theta}^0/Pr)/\ell$ with $v_{\theta}^0$ the maximum azimuthal velocity at the base of the stratified vortex cap (see values in table \ref{table}). An approximate expression for $\alpha(Ro,\mathcal{N}/f)$ can be derived from the cyclo-geostrophic and hydrostatic equations \eqref{geo3}, which are slightly more relevant in our case of small but finite Rossby number (see figure \ref{fig3}B-C) than the thermal wind balance \eqref{geo}. 
%
%
The relationship between $\alpha$, $Ro$ and $\mathcal{N}/f$ arises from the requirement that the pressure anomaly in the core related to the cyclo-geostrophic equation \eqref{geo31} must be the same as the pressure anomaly due to the positive buoyancy anomaly of the stratified vortex cap (cf. equation \eqref{geo32}). This is a type of consistency condition that leads to an expression for $\alpha(Ro,\mathcal{N}/f)$ that depends on the radial profile of azimuthal velocity and on the vertical profile of buoyancy. The formula for $\alpha(Ro,\mathcal{N}/f)$ was previously derived for vortices in fully-stratified fluids \cite{Hassanzadeh2012} and  lenticular vortices at the ocean surface \cite{DelaRosaZambrano2017}, and here we derive it for turbulent LSVs penetrating in a stably-stratified fluid. We find that the radial profile of LSVs in DNS matches reasonably well with the radial profile of shielded monopoles \cite{carton1989}, and that the vertical profile of buoyancy anomaly is well approximated by a constant substratified bottom {(i.e. a stably-stratified fluid region but with a stratification strength smaller than $N$, which is the stratification strength in the above stably-stratified fluid bulk)} with an exponentially-decaying cap (cf. appendix \ref{appA}). This yields the formula
\ba{}\label{alpha2}
\alpha = a_1\f{f}{\mathcal{N}}\sqrt{Ro(1+a_2Ro)},
\ea
with $a_1$ and $a_2$ parameters of order unity, given by
\ba{}\label{coeffs0}
a_1^2 =\f{\Gamma\lp\f{2}{\mu}\rp\mu^{\f{2}{\mu}-1}e^{\f{1}{\mu}}}{\f{b_0}{N^2h}\lp\f{\delta}{h}+1\rp +\lp\f{N_0}{N}\rp^2\f{\delta}{h}\lp\f{\delta}{2h}+1\rp}, \quad a_2 = \lp\f{e}{4}\rp^{1/\mu},
\ea
with $\Gamma$ the Gamma function, $\mu$ the steepness parameter of the velocity profile in $r$, $b'_0$ the buoyancy anomaly at the base of the stratified vortex cap and $N_0<N$ the stratification strength of the stratified vortex cap inside the convective layer  (derivation details are provided in appendix \ref{appA}). 

Equation \eqref{alpha2} is derived under the assumption that the LSV is in an infinitely deep and wide stably-stratified fluid, i.e. such that the stratified vortex cap doesn't reach the top boundary. In our simulations, we have both vertically confined and unconfined LSVs and our numerical box has a finite horizontal extent, such that equation \eqref{alpha2} cannot be expected to be satisfied exactly. Nevertheless, we show in figure \ref{fig4}A that the aspect ratio $h/\ell$ measured directly from the velocity profile $v_{\theta}$ in DNS matches very well with the theoretical prediction \eqref{alpha2} based on the problem parameters, such that the formula is applicable for both unconfined and confined LSVs.

From equation \eqref{alpha2} we can obtain an approximate expression for the maximum diameter of LSVs penetrating in a stably-stratified fluid. The maximum diameter of LSVs is the diameter of LSVs that are confined vertically and saturate naturally by boundary friction (in the absence of other saturating mechanisms). We show in figure \ref{fig4}B the radius of LSVs $\ell$ in our simulations as a function of $H/\alpha$ with $\alpha$ given by equation \eqref{alpha2}. To the left of the diagram, i.e. where $H$ is small, we have the results of LSVs that are confined vertically and saturate naturally. For such LSVs, we find that $\ell \approx H/(2\alpha)=\ell_{max}$, which we therefore define as the maximum radius of LSVs. To the right of the diagram, where $H$ is large and LSVs saturate horizontally at the box size before they reach the top boundary, we find that $\ell<\ell_{max}$, as expected. Note that based on figure \ref{fig4}B $\ell$ is in fact close to $\ell_{max}+\epsilon$ with $\epsilon\approx 0.15$ a small correction in the limit $H/\alpha \rightarrow 0$, which may be due to complicated boundary layer effects that are neglected in the present work.

We find that $\mu\approx 1$ such that $a_2\approx 2/3$ in all simulations, i.e. for both vertically confined and unconfined LSVs, and that $a_1\approx 2$ for unconfined LSVs but varies with the problem parameters, i.e. $a_1\in[2,6]$, for confined LSVs (cf. appendix \ref{appA}). Thus, in section \S\ref{sec:disc} we will use for the penetration depth of unconfined LSVs and for the maximum radius of confined LSVs the approximate formulas
\ba{}\label{pred}
h = 2\ell\alpha_0, \quad \ell_{max} = \f{H}{4\alpha_0}, \quad \text{with} \quad \alpha_0= \f{f}{\mathcal{N}}\sqrt{Ro\lp 1+\f{2Ro}{3}\rp},
\ea 
i.e. with $\ell_{max}$ an upper bound based on our DNS results (i.e. using the minimum of $a_1$).

\section{Discussion}\label{sec:disc}

{Our generic model relies on a minimum number of physical ingredients and parameters to reproduce self-consistently the two-layer dynamics of atmospheres, oceans, planetary cores and stars, which have a turbulent layer next to a stably-stratified one. Thus, while our study discards many details specific to the dynamics of atmospheres, oceans, planetary cores and stars, it captures a leading-order effect such that we suggest that the shape of LSVs obtained in our DNS, which consists of a depth-invariant cylinder in the turbulent layer and of a penetrating half dome in the stratified layer, is generic. We summarize our findings and discuss applications in the next paragraphs.} 

The LSVs are depth-invariant in the convective layer and decay in the stratified layer by thermal wind balance because the LSV core is positively buoyant. The growth of LSVs stops when LSVs penetrate through the entire stratified layer depth and reach the top no-slip boundary{, which dissipates the LSV energy by friction}. Thus, in addition to the {well-known beta-effect that saturates LSVs at the Rhines scale \cite[e.g.][]{rhines1975,vallis1993,Vasavada2005}} and to the presence of strong magnetic field \cite{Maffei2019}, boundary friction across a stably stratified layer constitutes a physically relevant saturation mechanism to quench the inverse cascade of rapidly rotating, convective turbulence in natural systems. {We note that wave radiation by geostrophic adjustment provides another mechanism for dissipating energy from LSVs. However, we haven't found LSVs saturating naturally without being confined vertically (and thus experiencing boundary friction), such that wave radiation is unlikely to be the main cause of LSV energy dissipation in our simulations.}

\begin{figure}[ht]
\centering
\includegraphics[width=0.6\linewidth]{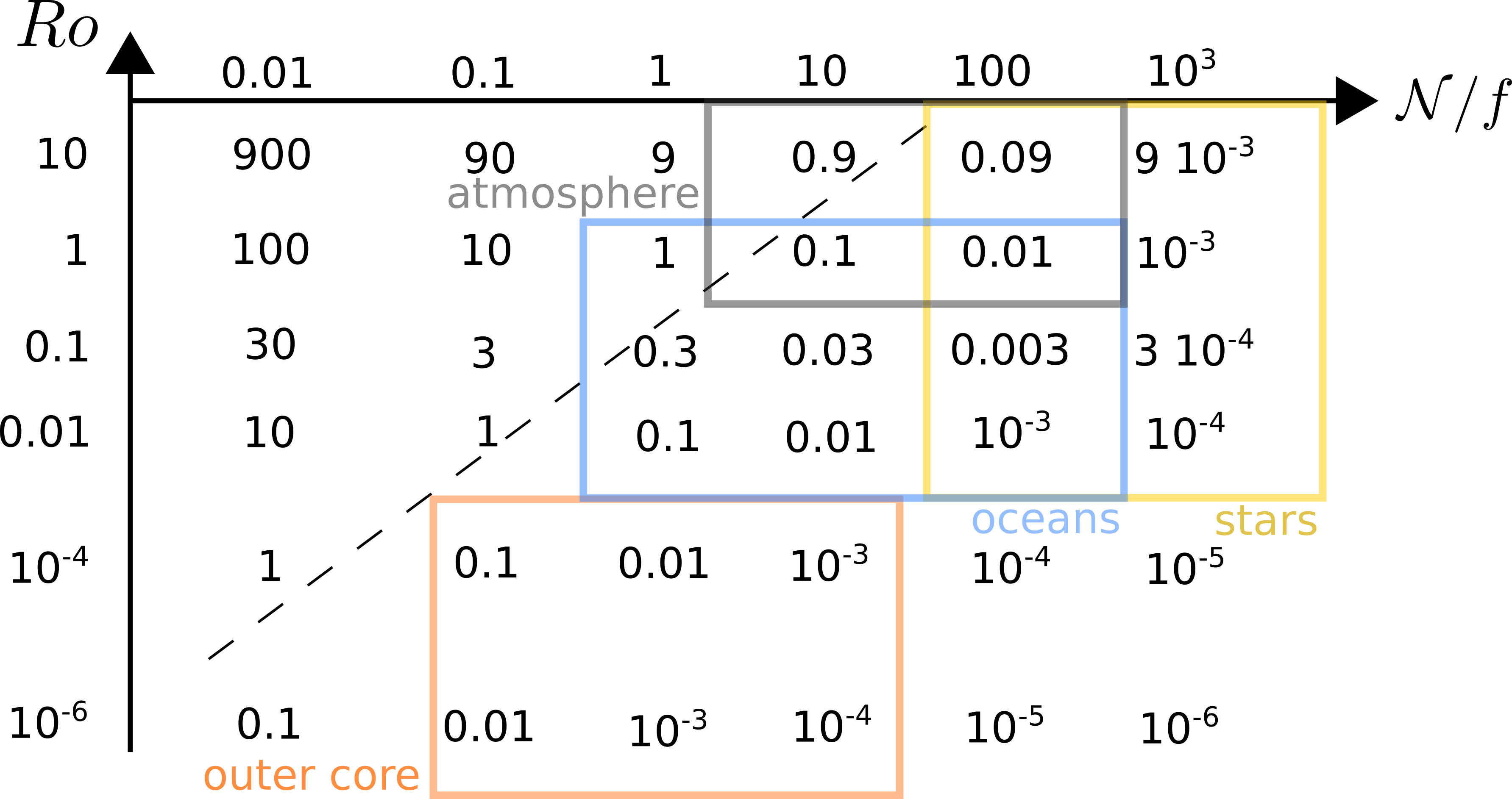}
\caption{Reference aspect ratio $\alpha_0$ of equation \eqref{pred} in $(\mathcal{N}/f,Ro)$ space. The colored rectangles highlight regions of the  $(\mathcal{N}/f,Ro)$ plane relevant to Earth's atmosphere (grey), oceans (blue), outer core (orange) and stars (yellow). $\alpha_0\geq 1$ above the dashed line (tall, penetrating LSVs), whereas $\alpha_0\leq 1$ below the dashed line (wide, weakly-penetrating LSVs).}
\label{fig5}
\end{figure}

LSVs studied in this work may be considered a simplified model of cyclones in Earth's atmosphere \cite{Emanuel2011}, and in particular of warm-core tropospheric cyclones penetrating into the stratosphere \cite{Thorpe1986}, of eddies in Earth's oceans \cite{Siegel2001,Bashmachnikov2015}, and of LSVs in Earth's outer core \cite{guervilly2019} and stars. Earth's atmosphere, oceans, outer core and stars have different fluid properties, such that the aspect ratio of the stratified cap of LSVs, and the penetration depth, depend on the geophysical or astrophysical fluid of interest. We give in figure \ref{fig5} different values of the aspect ratio $\alpha_0$ of equations \eqref{pred} in $(\mathcal{N}/f,Ro)$ plane, and we highlight regions relevant to Earth's atmosphere, oceans, outer core and stars. The atmosphere and oceans are relatively strongly stratified (i.e. $\mathcal{N}/f\geq 1$) and have high $Ro$ and moderate $Ro$, respectively. On the other hand, Earth's outer core is expected to be moderately stratified with small $Ro$, and stars have strong stratification (i.e. large buoyancy frequency compared to the rotation frequency) and moderate $Ro$. As a result, LSVs are expected to be wide and weakly-penetrating in Earth's outer core and stars, while moderately-penetrating in Earth's atmosphere and oceans.

For LSVs in Earth's atmosphere, if we take $Ro\sim 1$ and $\ell\sim 100$km, equation \eqref{pred} yields $h\in[2,20]$km for $\mathcal{N}/f\in[10,100]$. Thus, our model predicts that atmospheric LSVs can reach far into the stratosphere, and potentially all the way to the ozone layer found at $\approx 20$km when the turbulent planetary boundary layer is deep and atmospheric stability is low. In Earth's oceans, mesoscale eddies have typically $\ell\sim 100$km and $Ro\sim 10^{-2}$  (based on rms velocity $\sim 10$cm/s) \cite{klocher2013}, and submesoscale eddies have typically $\ell\sim 10$km and $Ro\sim 1$ \cite{badin2011}. Thus, the penetration depth of both eddy types is the same, i.e. $h\in[0.2,20]$km for $\mathcal{N}/f\in[1,100]$, which shows that surface eddies can penetrate relatively far into the thermocline and potentially reach the seabed, especially in weakly-stratified waters on the continental shelf. In the Earth's core, if we take $\ell \sim 30 $km and $Ro\sim 10^{-6}$ for the diameter and Rossby number of the most intense LSVs, as suggested in a recent study \cite{guervilly2019}, we find $h \sim 2$m for $\mathcal{N}/f=1$, which is a typical value used in previous works, e.g. \cite{Buffett2014}. This result suggests that upwellings and downwellings inside dominant LSVs in Earth's core do not promote exchanges of chemical species between the convection zone and far into the stably-stratified layer, unlike LSVs in the atmosphere and oceans. We note that non-dominant LSVs in Earth's core may penetrate farther into the stably-stratified layer. Previous works on fully-turbulent outer core dynamics studied LSVs at both planetary scale $\ell\sim 1000$km with $Ro\sim 10^{-5}$ and smaller scales $\ell \sim 100$km with $Ro\sim 10^{-4}$ \cite{Finlay2010,Gillet2015}. For such LSVs and $\mathcal{N}/f=1$, we find $h\sim 8$km and $h\sim 2$km, respectively. In stars, $Ro\sim 1$ is relevant for supergranulation \cite{Featherstone2016} and $\mathcal{N}/f\sim 10^3$ is a reasonable estimate for the stratification of the Sun \cite{Alvan2014}. If $\ell \sim 0.1R_{*}$ with $R_*$ the star radius, then $h\sim  0.8\;10^{-4}R_*$. Thus, LSVs in stars similar to the Sun are weakly-penetrating and cannot go through the tachocline, which is on the order of one percent of the stellar radius for the Sun \cite{Christensen2011}.

In Earth's oceans the thermocline shields LSVs from the seabed, and in Earth's outer core stably-stratified layers may shield LSVs at both the inner-core and core-mantle boundaries. The seabed and solid boundaries around Earth's outer core provide friction, which may play a role in the saturation of LSVs. The thickness of the stratified layer that starts at the thermocline of Earth's oceans and extends to the seabed is on the order of a few km, $H\in[1,10]$km, which means that the maximum diameter of LSVs in Earth's oceans, for a moderate stratification of $\mathcal{N}/f=10$, is $\ell_{max}\in[25,250]$km for mesoscale eddies ($Ro=10^{-2}$) and $\ell_{max}\in[2.5,25]$ for submesoscale eddies ($Ro=1$; cf. equation \eqref{pred}). Since the lower bound of $\ell_{max}$ lies in the range of observed ocean eddies, our work predicts that the seabed may play a role in limiting the size of ocean LSVs. The thickness of the stably-stratified layers in Earth's core is poorly constrained. Recent studies use $H \sim 100$km or more \cite{Hirose2013a,Yan2018}. For $H \sim 100$km, we find, for $\mathcal{N}/f\sim 1$, $\ell_{max}=25000$km for $Ro=10^{-6}$ and $\ell_{max}=2500$km for $Ro=10^{-4}$. The predicted $\ell_{max}$ is larger than the radial extent of Earth's outer core for both low and high $Ro$, such that our Cartesian approach is not valid anymore; for such large length scales, spherical geometry and the $\beta$ effect must be considered. Nevertheless, the large $\ell_{max}$ suggests that boundary friction is unlikely to be the relevant saturation mechanism for LSVs inside the Earth and that studies discarding the stably-stratified layer may use a stress-free boundary condition for the turbulent flows instead of a no-slip condition.

{It is worth mentioning that our simulations consider relatively low Rayleigh numbers. Specifically, $Ra=2\times 10^8\sim 5Ra_c$ where $Ra_c = 8.7 Ek^{-4/3}$ is the critical Rayleigh number in the limit $Ek\rightarrow 0$ \cite{kunnen2016}. This implies that the Reynolds number, $Re$, while large ($Re\sim O(10^3)$, see table \ref{table}), remains much smaller than the typical Reynolds number of flows in nature. Future studies should push to higher $Ra$ in order to challenge our conclusions in the limit of large $Re$. Instabilities of the LSV would also be worth investigating, as well as wave radiation, which may play a more important role at higher $Re$.}

Our model discards several physical effects, such as compressibility effects (which may be important for LSVs in the atmosphere, outer core and stars; cf. \cite{Mantere2011,Chan2013}), radiative transfers (atmosphere and stars), moist dynamics (atmosphere) and magnetic field (outer core and stars). Our work also neglects the spherical geometry, $\beta$ effects and the dynamics of the Ekman layer at the top of the stably-stratified fluid, which may affect the prediction for the shape and size of vertically-confined LSVs, as suggested by the variability of $a_1$ in equation \eqref{alpha2}. {The dynamic of the Ekman layer would be worth exploring in the limit $Re\rightarrow\infty$. In particular, it would be interesting to investigate whether the effect of the no-slip boundary conditions, which provide friction at large scales, completely disappear when $Re\rightarrow\infty$. Recent high-resolution studies indicate that free-slip and no-slip results diverge even in the canonical rotating Rayleigh-B\'enard experiment and at low Ekman numbers ($Ek$ down to $10^{-7}$) \cite{Stellmach2014,kunnen2016,Plumley2017}. This means that a departure between free-slip and no-slip results, as obtained in the present study, may be expected even in some practical situation due to Ekman pumping in no-slip cases.} {Here we have used Dirichlet boundary conditions for the temperature for simplicity and because of the close connection of our work with the well-known Rayleigh-B\'enard problem, which has been primarily studied experimentally, numerically and theoretically with fixed temperature boundary conditions. The effect of fixed-flux Neuman boundary conditions, which are arguably more relevant to conditions found in nature, would be worth exploring; however, we do not expect significant changes.} Future investigations taking into consideration one or several of the effects neglected in this study will help further our understanding of LSVs in nature.

\section*{Acknowledgments} 
The authors acknowledge funding by the European Research Council under the
European Union's Horizon 2020 research and innovation program through Grant no.
681835-FLUDYCO-ERC-2015-CoG. LAC acknowledges funding from the European Union's Horizon 2020
research and innovation programme under the Marie Sklodowska-Curie grant agreement No 793450. DL is supported by a PCTS fellowship and
a Lyman Spitzer Jr fellowship. Computations were conducted with support by the
HPC resources of GENCI-IDRIS (Grant no. A0020407543 and A0040407543) and
by the NASA High End Computing (HEC) Program through the NASA Advanced
Supercomputing (NAS) Division at Ames Research Center on Pleiades with allocations
GID s1647 and s1439.

\appendix

\section{Derivation of the aspect ratio of the stratified vortex cap}\label{appA}

We derive an expression for $\alpha$, i.e. the aspect ratio of the stratified vortex cap, by requiring that the pressure anomaly in the vortex core due to the cyclonic flow is the same as the pressure anomaly due to the buoyancy anomaly relative to the reference background or far-field value (cf. equation \eqref{geo2}) \cite{Hassanzadeh2012}. Here we define the vortex core as $r=0$ and $z=\xi_0=z(r=0,\overline{T}=0)$, i.e. where the isopycnal of maximum density intersects the LSV axis of rotation, and we recall that we denote by $\overline{p}',\overline{b}'$ and $\overline{p}_{\infty}(z),\overline{b}_{\infty}(z)$ the anomalous and far-field values, respectively, of the pressure and buoyancy fields, i.e. such that $\overline{p}=\overline{p}_{\infty}+\overline{p}'$ and $\overline{b}=\overline{b}_{\infty}+\overline{b}'$. 

We find that the radial profiles of $\overline{v_{\theta}}$ in our simulations match reasonably well with the generic radial profile of shielded monopoles \cite{carton1989} in  both the convective and stably-stratified layers, i.e. 
\ba{}\label{vapprox}
\overline{v_{\theta}}(r,z) \approx \f{v_{\theta}^{0}r e^{1/\mu}}{\ell}e^{-\f{1}{\mu}\lp\f{r}{\ell}\rp^{\mu}}
\ea
with $v_{\theta}^{0}(z)$ the maximum azimuthal velocity, $\ell(z)$ the radius where the velocity is maximum, and $\mu(z)\sim O(1)$ the best-fit profile steepness. We measure $v_{\theta}^{0}$ and $\ell$ from the DNS results at each $z$ and obtain $\mu(z)$ by least-square fit for $r\in[0,r_{max}]$. We show $\mu$ in figure 4 in SI \cite[][]{suppinfo}: $\mu$ is roughly equal to unity in the convection zone (exponential decay of vorticity in $r$) and then increases to approximately two in the stratified fluid (Gaussian decay, which is typical of vortices in stably-stratified fluids, e.g. \cite{Hassanzadeh2012}). We find that there is some variability of $\mu$ depending on $r_{max}$ (i.e. the extent over which we perform the best fit), which is not surprising since in our simulations the LSVs cannot relax to infinity but are instead horizontally periodic. We take the average of the best-fit values for $\mu$ for $r_{max}\in[1,1.5]$. We show in figure  \ref{gaussian} the normalized velocity $Ek\overline{v_{\theta}}(r)/Pr$ obtained in DNS (solid blue lines) as well as the proposed fit \eqref{vapprox} (blue circles) at different heights. The agreement is excellent in the  LSV core, i.e. solid lines and filled circles overlap very well for small $r$, and deteriorates slightly toward the outside, which is not surprising since the finite size and horizontal periodicity of the numerical domain prevents the velocity profile from relaxing to 0 far away.

A fit for the buoyancy anomaly $\overline{b}'$ is necessary to derive $\alpha$ but first requires to decompose $\overline{p}$ and $\overline{b}$ into far-field and anomalous values. Here, we require $\overline{p}_{\infty}$ and $\overline{b}_{\infty}$, which satisfy $\p_z\overline{p}_{\infty}=\overline{b}_{\infty}$, to be piecewise second- and first-order polynomials, since the buoyancy must have a purely diffusive profile far from the LSV. Thus, we seek far-field profiles of the form
\bsa{inffit}\label{pinffit}
& \overline{p}_{\infty} = c_0+\lb c_1(z-z_{\infty})+c_2(z-z_{\infty})^2 \rb \mathsf{H}(z_{\infty}-z)+\lb c_3(z-z_{\infty})+c_4(z-z_{\infty})^2\rb \mathsf{H}(z-z_{\infty}), \\ \label{binffit}
& \overline{b}_{\infty} = \lb c_1+2c_2(z-z_{\infty}) \rb \mathsf{H}(z_{\infty}-z)+\lb c_3+2c_4(z-z_{\infty})\rb \mathsf{H}(z-z_{\infty}),
\esa
where $\mathsf{H}$ is the Heaviside function, $c_i$ ($i=0,1,2,3,4$) are constants and $z_{\infty}\approx 1$ is the position of the convective-stratified interface at $r=\infty$ (which we let arbitrary, i.e. obtained by best-fit, although setting $z_{\infty}=1$, i.e. assuming no LSVs at infinity, results in minor changes). Let us denote by $P'$ the fit of pressure anomaly obtained by integrating equation \eqref{geo31} in $r$ with $\overline{v_{\theta}}$ substituted by \eqref{vapprox} and with the condition $P'=0$ at $r=z=\infty$, i.e. under the assumption of an unbounded fluid domain, which in practice translates to setting $P'=0$ at the maximum radius. The far-field parameters $c_i$ ($i=0,1,2,3,4$) and $z_{\infty}$ are obtained from a best-fit of $\overline{p}-P'$ with the proposed profile \eqref{pinffit}, and the anomalous pressure is then deduced as $\overline{p}'=\overline{p}-\overline{p}_{\infty}$. We show $\overline{p}'$ and the fit $P'$ by solid brown lines and circles, respectively, in figure \ref{gaussian}. The agreement between  $\overline{p}'$ in DNS and $P'$ is excellent in the LSV core though deteriorates toward the domain edge, i.e. at large $r$, as is the case for velocity (cf. blue lines and circles in figure \ref{gaussian}). The far-field buoyancy $\overline{b}_{\infty}$ is obtained from equation \eqref{binffit} and the buoyancy anomaly is finally deduced from $\overline{b}'=\overline{b}-\overline{b}_{\infty}$. We show $\overline{p}-\overline{p}'$ and $\overline{p}_{\infty}$, as well as $\overline{b}-\overline{b}'$ and $\overline{b}_{\infty}$ in figure 5 in SI \cite[][]{suppinfo}.

In our simulations, the buoyancy anomaly is well approximated along the rotation axis by a profile of the form
\ba{}\label{bapprox} 
\overline{b}'(r=0,z) \approx  \lcb \begin{array}{c}
b'_0+(z-\xi_0)N_0^2, \quad \xi_0\leq z<z_{\infty}  , \\
\lb b'_0+(z_{\infty}-\xi_0) N_0^2 \rb e^{-\f{z-z_{\infty}}{h}}, \quad z_{\infty}\leq z.
\end{array}   \right.
\ea
with $b'_0$ the buoyancy anomaly at $z=\xi_0$, $N_0$ the density restratification due to the LSV in $[\xi_0,z_{\infty}]$, $z_{\infty}\approx 1$ the profile transition height, and $h$ the overshooting depth parameter; $b'_0$ and $N_0$ are estimated from the simulation results, while $h$ is obtained by best fit for $z\geq \xi_0$. We denote $\delta=z_{\infty}-\xi_0$ the depth of restratification of the fluid below $z_{\infty}\approx 1$ (as shown in figure \ref{fig3}A). We show in figure  \ref{zprof} the buoyancy anomaly $b'(0,z)$ obtained in DNS (solid blue lines) as well as the proposed fit \eqref{bapprox} from the base of the stratified vortex cap upward (blue circles). The agreement between DNS results and the fitted profiles is reasonably good, i.e. solid lines and filled circles overlap well in the stratified LSV core (note that the fit can be improved by substituting $z_{\infty}$ with anoter free parameter). For completeness we also show the normalized pressure $\overline{p}'/(PrRa)$ obtained in DNS and the fit $P'$ at $r=0$ by solid red lines and red circles, respectively. Again, the DNS profiles and the proposed fits overlap well.


\begin{figure}
\centering
\includegraphics[width=0.8\textwidth]{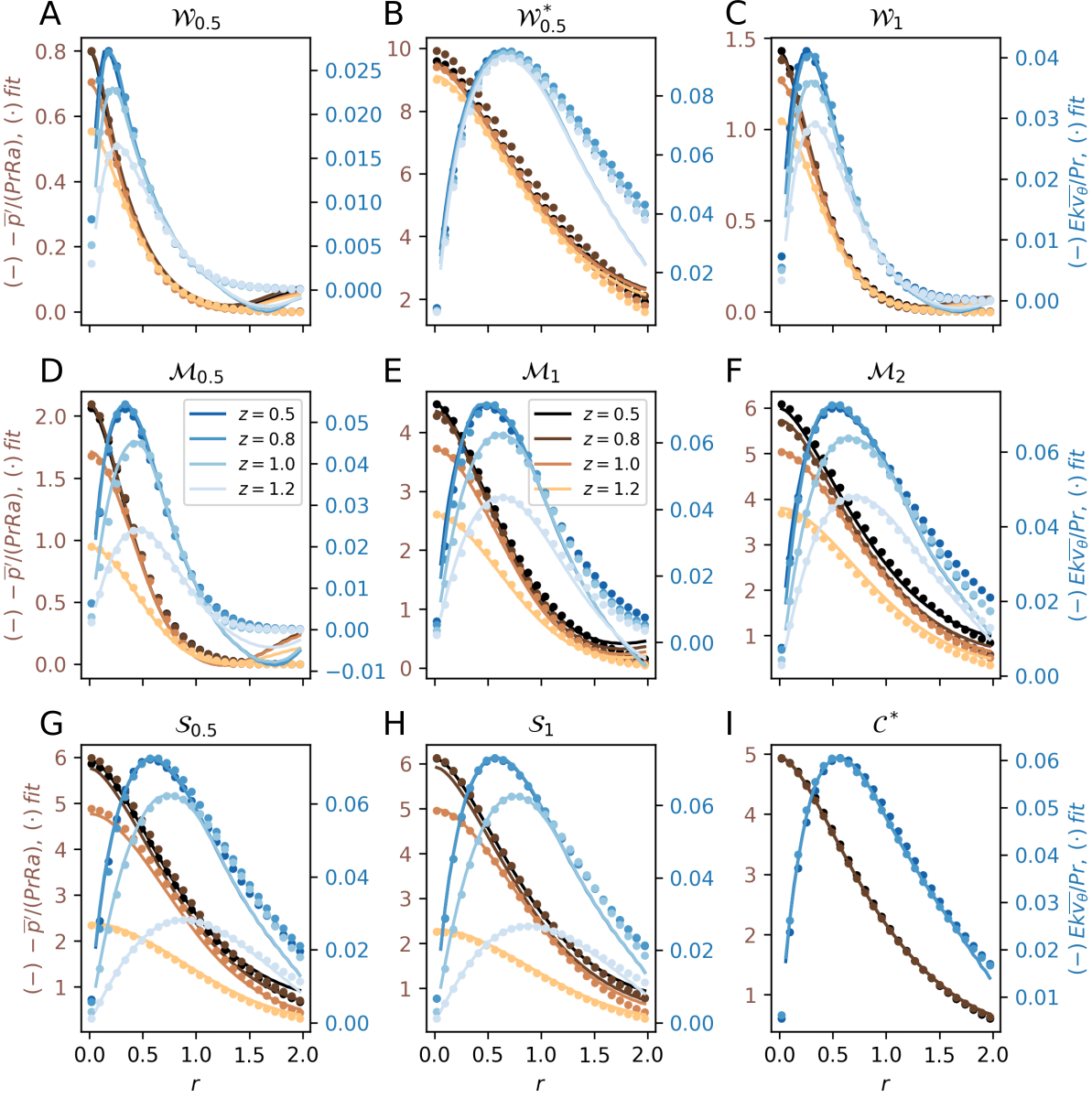}
\caption{Plots of pressure anomaly $\overline{p}'$ (brown solid lines) and normalized velocity $Ek\overline{v_{\theta}}/Pr$ (blue solid lines) along with the shielded-monopole fit proposed in equation \eqref{vapprox} (filled circles) for $z=0.5,0.8,1.0,1.2$ as functions of $r$. The results are shown for all simulations of table \ref{table} featuring LSVs.}\label{gaussian}
\end{figure}

\begin{figure}
\centering
\includegraphics[width=0.75\textwidth]{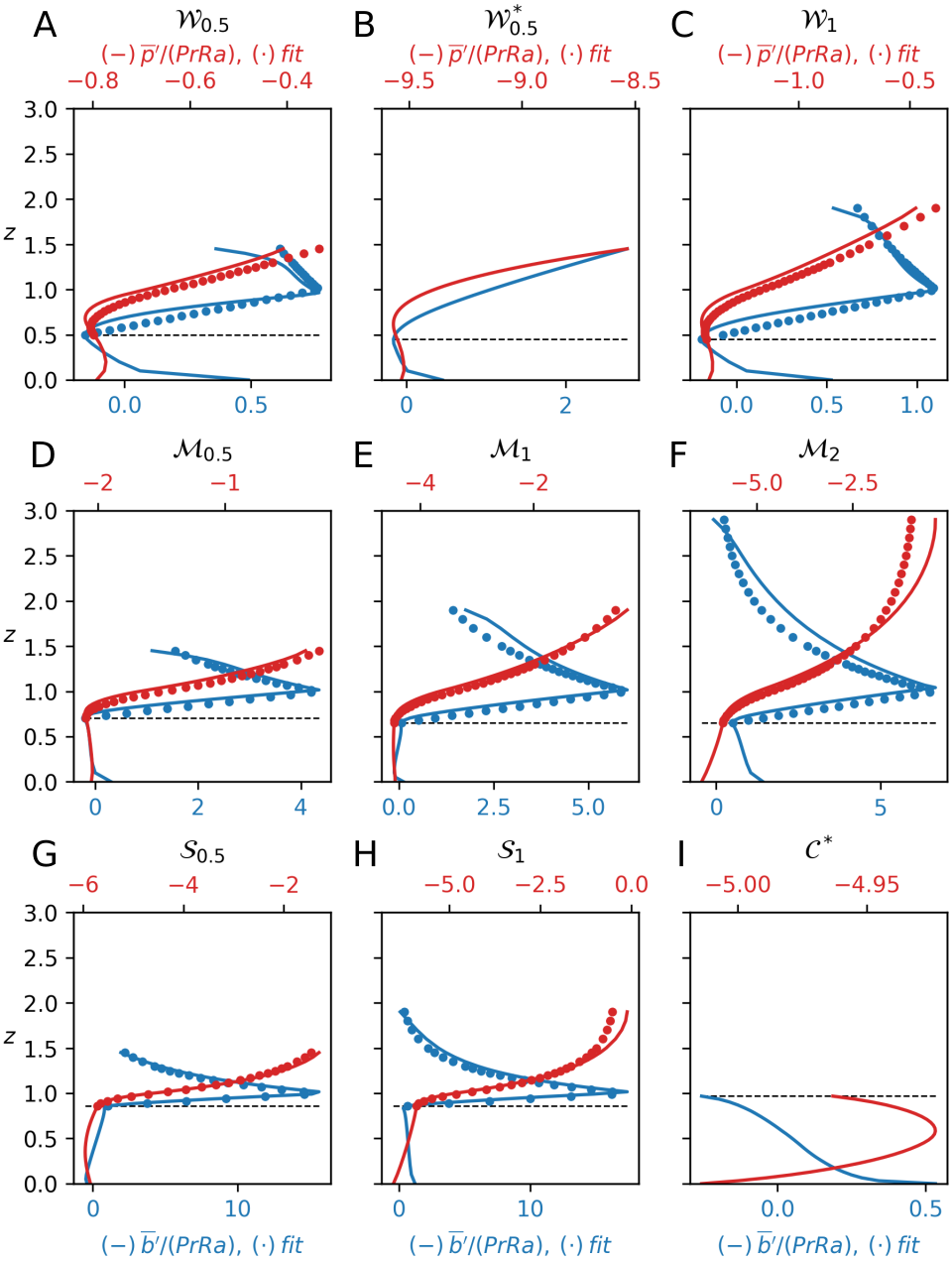}
\caption{Buoyancy $\overline{b}'$ (blue) and pressure $\overline{p}'$ (red) anomaly profiles with $z$ in the LSV core, i.e. at $r=0$ (solid lines). The filled circles show the linear profile followed by exponential relaxation fit for $\overline{b}'$ and corresponding fit for $\overline{p}'$ within the stratified vortex cap as expressed by equation \eqref{bapprox}. The dashed line shows $\xi_0=z(\overline{T}(r=0)=0)$, i.e. the lowest point of the stratified vortex cap. The results are shown for all simulations of table \ref{table} featuring LSVs, except for the fitted profiles, which are shown only for simulations with a stratified layer and a no-slip top boundary, for which our theory applies.}\label{zprof}
\end{figure}


We now derive a prediction for $\alpha$ as a function of the problem parameters using the proposed fits \eqref{vapprox} and \eqref{bapprox} for the velocity and buoyancy variables. Integrating \eqref{geo31} with $\overline{v_{\theta}}$ substituted by \eqref{vapprox} from the base of the stratified vortex cap ($0,\xi_0$) to ($\infty,\xi_0$) yields
\ba{}\label{vint}
\int_0^{\infty} \p_r p' dr = p'(\infty,\xi_0)-p'(0,\xi_0) = \int_0^{\infty} \lp\f{Prv_{\theta}}{Ek}+\f{v_{\theta}^2}{r}\rp dr = \ell^2\Gamma\lp\f{2}{\mu}\rp\mu^{\f{2}{\mu}-1}e^{\f{1}{\mu}} \f{Pr^2}{Ek^2} Ro\lb 1+\lp\f{e}{4}\rp^{1/\mu} Ro\rb
\ea
with $\Gamma(\cdot)$ the Gamma function and $Ro=Ek(v_{\theta}^0/Pr\ell)$ the Rossby number based on $v_{\theta}^0$ and $\ell$ at the base of the stratified layer ($Pr$ appears because we use the thermal diffusive time scale for normalization). Integrating \eqref{geo32} vertically along $r=0$  with $\overline{b}'$ substituted by \eqref{bapprox} then yields
\ba{}\label{bint}
\int_{\xi_0}^{\infty} \p_z p' dz = p'(0,\infty)-p'(0,\xi_0) = \int_{\xi_0}^{\infty} b' dz = b_0(\delta+h)+N_0^2\delta\lp\f{\delta}{2}+h\rp.
\ea
Assuming that the pressure anomaly far from the vortex is 0, i.e. $p'(\infty,z_{\delta})=p'(0,\infty)=0$, and equating \eqref{vint} and \eqref{bint} we finally obtain for the aspect ratio squared 
\ba{}\label{alpha}
\alpha^2 = \f{h^2}{\ell^2} = \f{a_1^2Ro(1+a_2Ro)}{Ek^2N^2/Pr^2},
\ea
which yields equation \eqref{alpha2} with $EkN/Pr$ rewritten as $\mathcal{N}/f$ ($\mathcal{N}$ the dimensional buoyancy frequency), and we recall that the parameters $a_1$ and $a_2$, already presented in equation \eqref{coeffs0}, are given by
\ba{}\label{coeffs}
a_1^2 =\f{\Gamma\lp\f{2}{\mu}\rp\mu^{\f{2}{\mu}-1}e^{\f{1}{\mu}}}{\f{b_0}{\bar{N}^2h}\lp\f{\delta}{h}+1\rp +\lp\f{N_0}{\bar{N}}\rp^2\f{\delta}{h}\lp\f{\delta}{2h}+1\rp}, \quad a_2 = \lp\f{e}{4}\rp^{1/\mu}.
\ea
We find that $a_1$ is approximately constant, i.e. $a_1\approx 2$, for unconfined LSVs, while for confined LSVs $a_1\in[2,6]$ shows some variability (cf. figure \ref{scatter1}A). Furthermore we find that $a_2\approx 2/3$ in all simulations (figure \ref{scatter1}B). The variability of $a_1$ for confined LSVs comes from the fact that (i) the fit proposed for the buoyancy anomaly does not accomodate for the Ekman layer dynamics near the top boundary, and (ii) the buoyancy anomaly at the base of the stratified vortex cap ($b_0$) is sensitive to $EkN$ in the weak stratification limit (small $S$). The latter point can be seen in figure \ref{scatter2}C where $b_0/(N_0^2\delta)$ decreases with decreasing $EkN/Pr$, such that the vortex cap bottom is lighter than its surrounding for large stratification but is heavier for low stratification: this is a complicated effect which we do not attempt to predict but which may be expected to be negligible in the limit of strong stratification (i.e. with a sharper convective-stratified interface and reduced vertical exchanges \cite{Couston2017prf}). Two observations are worth noting: $\delta/h\approx 0.5$ is roughly constant, in particular in the limit of large stratification (cf. figure \ref{scatter2}A), and $N_0/\bar{N}\approx 0.7$ is roughly constant in all simulations (i.e. the stratification strength of the substratified bottom is always equal to roughly 0.7 times the background stratification; cf. figure \ref{scatter2}B). Both observations are consistent with the fact that $a_1$ should be roughly constant across simulations. Finally, $a_2\approx 2/3$ is constant because $\mu\approx 1$ at the base of the stratified vortex cap in all simulations. \\ \\

\begin{figure}[H]
\centering
\includegraphics[width=0.6\textwidth]{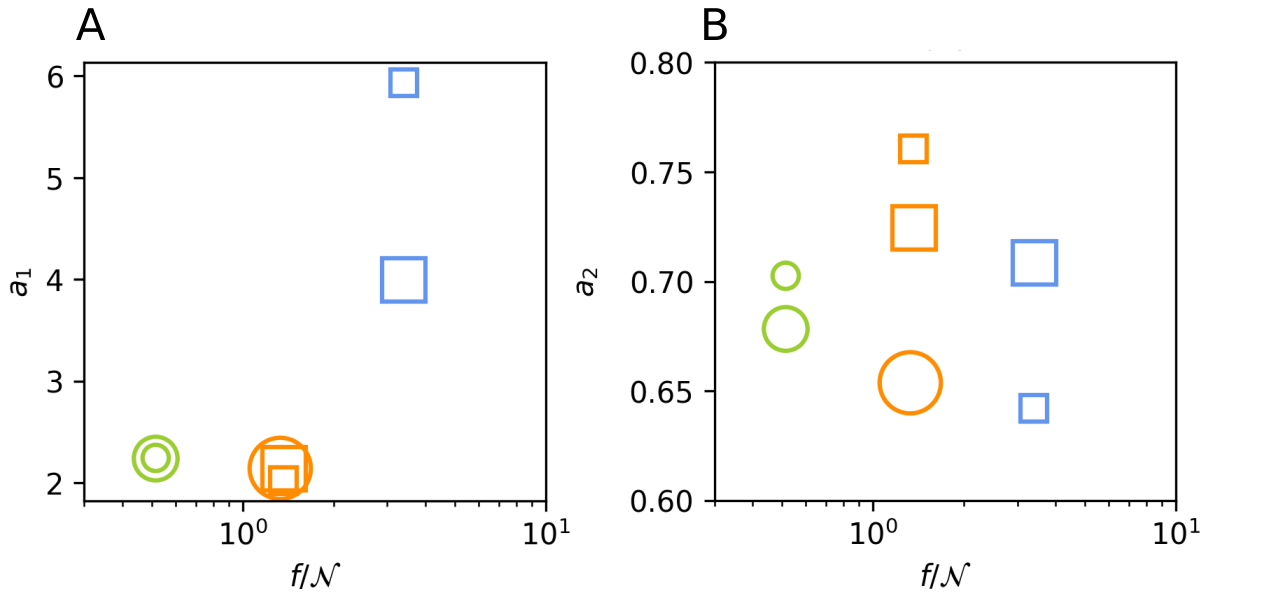}
\caption{Best-fit values for (A) $a_1$ and (B) $a_2$ of equation \eqref{coeffs}. The symbols refer to the same simulations as in figure \ref{fig4}.}\label{scatter1}
\end{figure} 

\begin{figure}[H]
\centering
\includegraphics[width=1\textwidth]{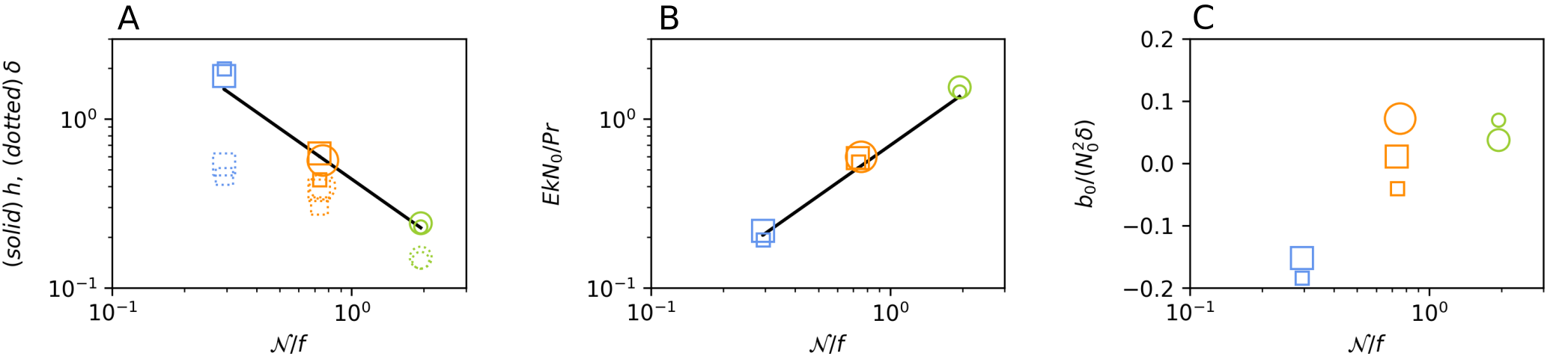}
\caption{Best-fit values for (A) $h$ (solid markers), $\delta$ (dotted markers), (B) $EkN_0/Pr$ and (C) $b_0/(N_0^2\delta)$ (cf. equations \eqref{coeffs}). The symbols refer to the same simulations as in figure \ref{fig4}.}\label{scatter2}
\end{figure}







\bibliography{rotation}

\end{document}